\documentclass[12pt, onecolumn]{IEEEtran}
\renewcommand{\baselinestretch}{1.5}

\usepackage{graphicx}

\newtheorem{theorem}{\bf Theorem}

\newtheorem{proposition}{\bf Proposition}

\newtheorem{definition}{\bf Definition}

\begin{document}
%
\title{A New Approach to Random Access: Reliable Communication and Reliable Collision Detection}
%
%
\author{Jie~Luo, Anthony Ephremides
\thanks{Jie Luo is with the Electrical and Computer Engineering Department, Colorado State University, Fort Collins, CO 80523. E-mail: rockey@engr.colostate.edu. Anthony Ephremides is with the Electrical and Computer Engineering Department, University of Maryland, College Park, MD 20742. E-mail: etony@umd.edu.}
\thanks{This work was supported by the National Science Foundation under Grants CCF-0728826 and CCF-0728966.}
}

\maketitle

\begin{abstract}
This paper applies Information Theoretic analysis to packet-based random multiple access communication systems. A new channel coding approach is proposed for coding within each data packet with built-in support for bursty traffic properties, such as message underflow, and for random access properties, such as packet collision detection. The coding approach does not require joint communication rate determination either among the transmitters or between the transmitters and the receiver. Its performance limitation is characterized by an achievable region defined in terms of communication rates, such that reliable packet recovery is supported for all rates inside the region and reliable collision detection is supported for all rates outside the region. For random access communication over a discrete-time memoryless channel, it is shown that the achievable rate region of the introduced coding approach equals the Shannon information rate region without a convex hull operation. Further connections between the achievable rate region and the Shannon information rate region are developed and explained.
\end{abstract}


%
\IEEEpeerreviewmaketitle


\section{Introduction}
\label{SectionI}
Classical information theory regards each transmitter in a multiuser communication system as backlogged with an infinite reservoir of traffic \cite{ref Shannon48}\cite{ref Cover05}. To achieve reliable communication, transmitters first \underline{jointly} determine their codebooks and their information rates and share this information with the receiver and with each other. The encoded symbols are then transmitted to the receiver continuously over a long time. Channel capacity and channel coding theorems are proved using the standard argument of jointly typical sequences by taking the sequence (or codeword) length to infinity \cite{ref Shannon48}\cite{ref Cover05}.

By allowing a small acceptable communication error probability, information theoretic results can be extended to channel coding within a finite-length time duration \cite{ref Gallager65}. Consequently, in a time-slotted communication model, if information bits arrive stochastically and queue at the transmitters, the latter can jointly adapt their information rates in each time slot to optimize certain system performance based on coding theoretic results and on the status of the message queues \cite{ref Telatar95}\cite{ref Berry02}\cite{ref Yeh02}. Determination of fundamental performance limitations, such as the throughput and queuing delay tradeoff, can therefore be obtained as in \cite{ref Berry02}. Although such an extension enabled a modest incorporation of stochastic traffic arrivals in information theoretic analysis, it inherited the key assumption, and hence also the limitation, of joint information rate determination among transmitters in each time slot \cite{ref Telatar95}\cite{ref Berry02}\cite{ref Yeh02}.

For various reasons, such as bursty traffic arrivals, timely data dissemination, and cognitive networking \cite{ref Zhao07}, transmitters and receivers in a communication network may not always want, or be able, to jointly design channel codes and determine communication rates. Random channel access is therefore commonly seen in practical networks \cite{ref Ephremides98}. In random access communication, transmitters make distributed channel access decisions, such as whether or not to transmit a packet. For example, if we regard the idling of a transmitter as setting its communication rate at zero and regard the transmission of a packet as setting the communication rate of a transmitter at a non-zero value, then communication rates of the transmitters are determined individually. The rate information is shared neither among the transmitters nor with the receiver. Distributed rate determination leads to unavoidable packet collision \cite{ref Bertsekas92}. When the joint rates of the transmitters are chosen such that reliable packet recovery is not possible, for efficient functioning of the upper layer protocols, the receiver is required to report a packet collision rather than blindly forward unreliable messages to the upper layers \cite{ref Bertsekas92}.

Due to the challenges that result from relaxing the joint rate determination assumption among transmitters and the receiver, and from making collision report decisions at the receiver without knowing the communication rates, information theoretic analysis has not been fully and successfully extended to practical random access systems. Consequently, without the support of rigorous coding theorems, standard networking practice often focuses on throughput optimization using packet-based channel models \cite{ref Luo06}. The explicit connection of the packet-based channel models to the physical layer channel is usually not specified except through the limited means of packet erasure channels. Networking practice allows bursty traffic arrivals and distributed determination of communication parameters. However, the use of packet-based communication models and the lack of rigorous coding theoretic analysis essentially prevent an insightful understanding of the impact of physical layer communication to upper layer networking \cite{ref Ephremides98}.

In this paper, we propose an approach that holds promise in extending information theoretic analysis to packet random access systems with bursty traffic. The essence of our approach consists of using the classical foundation of coding for \underline{each packet} and explicitly building-in the support of random access operations and bursty traffic phenomena in the following sense. In our coding approach, each transmitter determines its communication rate by choosing the number of data bits to encode in each packet. It requires neither joint communication rate determination among transmitters, nor pre-sharing communication rate information with the receiver. It also enables collision detection at the receiver whenever reliable packet recovery is not possible. Although defined quite differently from classical channel coding, we find that the introduced coding approach does lead to a meaningful achievable rate region characterization that is consistent with current understanding and methodology of Information Theory. More specifically, we define an achievable region on the communication rates such that reliable packet recovery is supported for all rates inside the region and reliable packet collision detection is supported for all rates outside the region\footnote{Note that communication rates are chosen arbitrarily and the rate information is unknown at the receiver.}. For random multiple access communication over a discrete-time memoryless channel using a class of random coding schemes, we show that the achievable rate region of the introduced coding approach equals the Shannon information rate region without a convex hull operation. Although we only illustrate our results in single-user and simple multiple access systems, the general problem formulation shown in the paper can be extended to other random access scenarios\footnote{We want to emphasize that our work does not purport to \underline{fully} bridge the gap between Networking and Information Theory. However, it does provide a useful link between rigorous communication rate determination and practical random access networking.}.

Next, we start with a detailed explanation of the coding approach in a single user system (i.e., single transmitter-receiver pair) in Section \ref{SectionII}. We then extend it to a random multiple access system and prove the main coding theorem in Section \ref{SectionIII}. Further extensions are discussed in Section \ref{SectionIV}.

\section{A New Packet Coding Approach -- The Single User Case}
\label{SectionII}

Let us first consider a single user communication system over a discrete-time memoryless channel. The channel is modeled by a conditional distribution function $P_{Y|X}$ where $X\in {\cal X}$ is the channel input symbol and $Y\in {\cal Y}$ is the channel output symbol. The sets ${\cal X}$ and ${\cal Y}$ are the (finite) input and output alphabets. We assume that time is slotted with each slot equaling $N$ symbol durations, which is also the length of a packet. Unless otherwise specified, we will confine our focus on block channel codes of length $N$ that represent coding \underline{within each packet}. Throughout this section, we assume that communication channel is time-invariant. The channel is known at the receiver but {\it unknown} at the transmitter. Our main objective is to use a relatively simple system model to introduce the basic coding approach that provides multiple rate options to the transmitter and enables collision detection at the receiver. Proofs of the claims and theorems given in this section are skipped since they are trivially implied by the more general theorems given in Section \ref{SectionIII}.

\subsection{Random Coding with Multiple Rate Options}
\label{SectionII.A}
Consider the simple case when the transmitter uses a classical random coding scheme, originally introduced in \cite{ref Shamai07}. The coding scheme is described as follows. Let ${\cal L}=\{{\cal C}_{\theta}: \theta \in \Theta\}$ be a library of codebooks, indexed by a set $\Theta$. Each codebook contains $2^{NR_0}$ codewords of length $N$, where $R_0$ is a predetermined rate parameter. Denote by $[{\cal C}_{\theta}(w)]_j$ the $j$th symbol of the codeword corresponding to message $w$ in codebook ${\cal C}_{\theta}$. Assume that, at the beginning of a time slot, the transmitter randomly generates a codebook index $\theta$ according to a distribution $\gamma$. The distribution $\gamma$ and the codebooks ${\cal C}_{\theta}$ are chosen such that the random variables $X_{w,j} : \theta \to [{\cal C}_{\theta}(w)]_j$, $\forall j, w$ are i.i.d. according to a predetermined input distribution $P_{X}$. We assume that the code library and the value of $\theta$ are both known at the receiver, that is, the receiver knows the randomly generated codebook. This can be achieved by sharing the random codebook generation algorithm with the transmitter. Based upon this information and upon the channel output, the receiver determines an estimate $\hat{w}$ of the transmitted message $w$. Define $P_e(w)=Pr\{\hat{w}\ne w\}$ as the decoding error probability given that $w$ is the transmitted message. By following the analysis in \cite{ref Shamai07}, it is easily seen that, if $R_0 < I(X; Y)$, there exists a sequence of decoding algorithms that achieve $\lim_{N\to \infty} P_e(w)=0$ for all $w$. The asymptotic result here should be interpreted as: given two small positive constants $\epsilon_1$, $\epsilon_2$, there exists a threshold $N(\epsilon_1, \epsilon_2)$, such that the conditions $N>N(\epsilon_1, \epsilon_2)$ and $R_0 < I(X; Y)-\epsilon_2$ imply $P_e(w)< \epsilon_1$ for all $w$ \cite{ref Polyanskiy10}. Although understanding of the tradeoff between $N$, $\epsilon_1$, $\epsilon_2$ is important, it is outside the scope of this paper. We will not repeat this well-known observation in the rest of the paper.

Now recall the standard practice of packet networking with bursty traffic \cite{ref Bertsekas92}. Depending on message availability, in each time slot, the transmitter will either stay idle or transmit a packet according to data availability and the MAC layer protocol. Suppose that the same coding scheme is used in multiple time slots, which means that, when the channel code is designed, the transmitter does not know whether or not a message will be available in a particular time slot. To model the idle operation, we regard ``idle" as a specific channel input symbol and insert a particular codeword ${\cal C}_{\theta}(1)=\{\mbox{idle}, \dots, \mbox{idle}\}$ into every codebook in the library ${\cal L}$. When no input data is available, we declare $w=1$ and the transmitter sends ${\cal C}_{\theta}(1)$ through the channel. It can be shown that, if $R_0< I(X; Y)$, we can still achieve $\lim_{N\to \infty} P_e(w)=0$ for all $w$.

Based on the above coding scheme, we will now introduce a communication rate parameter $r$. According to the usual practice, when the transmitter idles, we say the communication rate is $r=0$, otherwise, the communication rate is $r=R_0$. We say the codebook has two classes of codewords. The first class contains one codeword ${\cal C}_{\theta}(1)$ corresponding to $r=0$. The second class contains $2^{NR_0}$ codewords corresponding to $r=R_0$. The coding scheme enables the transmitter to choose its communication rate $r\in \{0, R_0\}$ without sharing the rate information with the receiver. If $R_0<I(X; Y)$, reliable recovery of the $(w, r)$ pair can be achieved asymptotically.

Next, let us consider a more complicated situation. Assume that each codebook in the library contains three classes of codewords. As before, the first two codeword classes contain one and $2^{NR_0}$ codewords, respectively. The third class contains $2^{NR_1}$ codewords corresponding to $r=R_1$. We assume that the index distribution $\gamma$ and the codebooks are designed such that random variables $X_{w,j} : \theta \to [{\cal C}_{\theta}(w)]_j$, $\forall j, w$ are independently distributed. Codeword symbols in the second class are i.i.d. according to an input distribution $P_{X|R_0}$, while codeword symbols in the third class are i.i.d. according to an input distribution $P_{X|R_1}$. Note that the two input distributions may be different. Let $I_{R_0}(X;Y)$ and $I_{R_1}(X;Y)$ be the mutual information between channel input and output symbols computed using input distributions $P_{X|R_0}$ and $P_{X|R_1}$, respectively. Assume that $R_0<I_{R_0}(X;Y)$, while $R_1>I_{R_1}(X;Y)$. In this case, the coding scheme provides three rate options $r\in \{0, R_0, R_1\}$ at the transmitter. However, it is clear that reliable message recovery is not always possible. Nevertheless, rather than trying to recover the messages blindly, let us assume that the receiver {\it intends} to achieve a different set of objectives. For all messages corresponding to $r=0$ and $r=R_0$, the receiver {\it intends} to recover the messages. For all messages corresponding to $r=R_1$, the receiver {\it intends} to report a packet ``collision"\footnote{Note that the term ``collision" is used to maintain consistency with the networking terminology. Throughout the paper, ``collision" means packet erasure, irrespective whether it is caused by multi-packet interference or not.}. Note that the receiver needs to achieve these objectives without knowing the actual communication rate $r$. To be more specific, let $(w, r)$ be the transmitted message and rate pair. The receiver either outputs an estimated pair $(\hat{w}, \hat{r})$, or outputs a ``collision". For $(w, r)$ with $r\in \{0, R_0\}$, we define $P_e(w, r)=Pr\{(\hat{w}, \hat{r})\ne (w, r)\}$ as the decoding error probability. For $(w, r)$ with $r=R_1$, we define $P_c(w, r)=Pr\{\mbox{collision}\}$ as the collision detection probability. It can be shown that, there exists a sequence of decoding algorithms to asymptotically achieve $\lim_{N\to\infty}P_e(w, r)=0$ for all $(w, r)$ with $r\in \{0, R_0\}$, and $\lim_{N\to\infty}P_c(w, r)=1$ for all $(w, r)$ with $r=R_1$.

In the above coding scheme, the transmitter has multiple rate options. The encoding scheme can be designed with only channel alphabet information. If the receiver knows the communication channel, whenever the transmitter sends a message with rate $r<I_r(X;Y)$, the receiver can asymptotically recover the message with an error probability arbitrarily close to zero; whenever the transmitter sends a message with rate $r>I_r(X;Y)$, the receiver can asymptotically report a collision with a probability arbitrarily close to one.

\subsection{Generalized Random Coding Scheme and The Standard Rate}
\label{SectionII.B}
In the previous section, we defined communication rate as the number of data bits per symbol encoded in a packet. Performance of the system was presented in terms of the asymptotic decoding and collision report probabilities corresponding to different rate values. However, in the following example, we show that, so long as the input distributions of the random coding scheme are given, the performance limitation of the system is actually independent of the rate options defined at the transmitter.

Consider the single user system where the same input distribution is used for all codewords in the random coding scheme. Suppose that the receiver chooses an arbitrary rate parameter $R$. Codewords are partitioned into two classes, with the first class containing the first $2^{NR}$ codewords and the second class containing the rest of the codewords. It can be shown that, so long as $R< I(X;Y)$, asymptotically, the receiver is able to decode reliably if the transmitted codeword is in the first class, and to report a collision reliably if the transmitted codeword is in the second class. Fundamental performance limitation of the random access system is characterized by $R<I(X;Y)$, which is determined by the input distribution of the random coding scheme. However, it is independent of the communication rate assignments and the overall size of the codebook.

More specifically, since the transmitter uses a codebook to map a message and rate pair $(w, r)$ into a codeword, one can regard $(w, r)$ as a ``macro" message, and index its possible values using a variable $W$. Communication rate then becomes a function of $W$, as defined by the transmitter. If we specify the random coding scheme using the message index $W$, as opposed to $(w, r)$, then performance limitation of the system can also be characterized as a function of $W$. From this viewpoint, the mapping $W \to (w, r)$, which associates practical variables with $W$, is actually independent from the performance limitation of the random access system.

To make the above argument rigorous, we introduce the concept of {\it standard rate} $r(W)$ as follows.

\begin{definition}{\label{Definition1}} ({\bf standard communication rate})
Assume that codebook ${\cal C}$ has $2^{NR_{\max}}$ codewords of length $N$, where $R_{\max}$ is a large finite constant. Let the corresponding messages or codewords be indexed by $W\in \{1, \dots, 2^{NR_{\max}}\}$. For each message $W$, we define its {\it standard communication rate}, in bits per symbol, as $r(W)=\frac{1}{N}\log_2 W$. $\QED$
\end{definition}

Note that, since the $r(W)=\frac{1}{N}\log_2 W$ function is invertible, the standard rate should be regarded as an equivalent form of the message index $W$. Practical meaning of the standard rate depends on how the transmitter maps $W$ to the message and communication rate pair. However, as explained above, we will detach such mapping from the performance analysis of the system.

Next, we introduce a generalized random coding scheme, in which, symbols of different codewords, as opposed to different codeword classes, can be generated according to different input distributions.

\begin{definition}{\label{Definition2}} ({\bf generalized random coding})
Let ${\cal L}=\{{\cal C}_{\theta}: \theta \in \Theta\}$ be a library of codebooks. Each codebook in the library contains $2^{NR_{\max}}$ codewords of length $N$, where $R_{\max}$ is a large finite constant. Let the codebooks be indexed by a set $\Theta$. Let the actual codebook chosen by the transmitter be ${\cal C}_{\theta}$ where the index $\theta$ is a random variable following distribution $\gamma$. We define $({\cal L}, \gamma)$ as a {\it generalized random coding scheme} following distribution $P_{X|W}$, given that the following two conditions are satisfied. First, the random variables $X_{W,j} : \theta \to [{\cal C}_{\theta}(W)]_j$, $\forall j, W$, are independent. Second, given $W$, $X_{W,j}$ are i.i.d. according to the input distribution $P_{X|W}$. $\QED$
\end{definition}

We now define a sequence of generalized random coding schemes that follows an asymptotic input distribution.

\begin{definition}{\label{Definition3}}
Let $\{({\cal L}^{(N)}, \gamma^{(N)} )\}$ be a sequence of random coding schemes, where $({\cal L}^{(N)}, \gamma^{(N)} )$ is a generalized random coding scheme with codeword length $N$ and input distribution $P^{(N)}_{X|W^{(N)}}$. Assume that each codebook in library ${\cal L}^{(N)}$ has $2^{NR_{\max}}$ codewords. Let $P_{X|r}$ be an input distribution defined as a function of the standard rate $r$, for all $r\in [0, R_{\max}]$. We say that $\{({\cal L}^{(N)}, \gamma^{(N)} )\}$ follows an asymptotic input distribution $P_{X|r}$, if for all $\{W^{(N)}\}$ sequences\footnote{Here $\{W^{(N)}\}$ refers to an arbitrary {\it deterministic}, i.e., not random, message sequence. Each message $W^{(N)}$ in the sequence takes value in the index set $\{1, \dots, 2^{NR_{\max}}\}$. The standard rate function $r(W^{(N)})$ is defined as $r(W^{(N)})=\frac{1}{N}\log_2 W^{(N)}$.} with well defined rate limit $\lim_{N\to\infty}r(W^{(N)})$, we have
\begin{equation}
\lim_{N\to \infty}P_{X|W^{(N)}}^{(N)}=\lim_{N\to \infty} P_{X| r(W^{(N)})}.
\label{DistributionConvergence}
\end{equation}
We assume that both $P_{X|W^{(N)}}^{(N)}$ and $P_{X| r(W^{(N)})}$ converge uniformly to their limits with respected to $r(W^{(N)})$. However, since we do not assume that $P_{X|r}$ should be continuous in $r$, we may not have $\lim_{N\to \infty} P_{X| r(W^{(N)})}=P_{X|\lim_{N\to\infty}r(W^{(N)})}$.
$\QED$
\end{definition}

\subsection{Achievable Rate Region of Single-user Random Access Communication}
\label{SectionII.C}
Consider single-user random access communication using a generalized random coding scheme $({\cal L}, \gamma )$ with codeword length $N$. In each time slot, the transmitter randomly generates a codebook index $\theta$ and uses codebook ${\cal C}_{\theta}$ in the library to map a message $W$ into a codeword. Assume that the channel and the codebook ${\cal C}_{\theta}$ are known at the receiver. Based on this information and the received channel symbols, the receiver either outputs a message estimate $\hat{W}$ or reports a collision. Let $P_{e|\theta}({\cal C}_{\theta}, W)=Pr\{\hat{W}\ne W\}$ be the probability, conditioned on $\theta$, that the receiver is not able to recover message $W$. Let  $P_{c|\theta}({\cal C}_{\theta}, W)=Pr\{\mbox{collision}\}$ be the conditional probability that the receiver reports a collision. We define $P_e(W)$ and $P_c(W)$ as the unconditional error probability and the unconditional collision probability of message $W$, respectively, as follows
\begin{equation}
P_e(W)=E_{\theta}[P_{e|\theta}({\cal C}_{\theta}, W)], \quad P_c(W)=E_{\theta}[P_{c|\theta}({\cal C}_{\theta}, W)].
\end{equation}

With this error probability definition, we define an achievable rate region of the system as follows.

\begin{definition}{\label{Definition4}}
Consider single user random access communication using a sequence of random coding schemes $\{({\cal L}^{(N)}, \gamma^{(N)} )\}$, where $({\cal L}^{(N)}, \gamma^{(N)} )$ is a generalized random coding scheme corresponding to codeword length $N$. Let $R\subseteq [0, R_{\max}]$ be a region of standard rates. Let $R_c$ be the closure of $R$. We say $R$ is {\it asymptotically achievable} if there exists a sequence of decoding algorithms under which the following two conditions are satisfied. First, for all message sequences $\{W^{(N)}\}$ with $r(W^{(N)})\in R$ for all $N$ and $\lim_{N\to\infty}r(W^{(N)})\in R$, where $r(W^{(N)})=\frac{1}{N}\log_2 W^{(N)}$ is the standard rate function, we have $\lim_{N\to\infty}P_e(W^{(N)})=0$. Second, for all message sequences $\{W^{(N)}\}$ with $r(W^{(N)})\not\in R_c$ for all $N$ and $\lim_{N\to\infty}r(W^{(N)})\not\in R_c$, we have $\lim_{N\to\infty}P_c(W^{(N)})=1$. $\QED$
\end{definition}

We are now ready to formally present the coding theorem for a single-user random access system.

\begin{theorem}{\label{Theorem1}}
Consider single user random access communication over a discrete time memoryless channel $P_{Y|X}$. Assume that the transmitter is equipped with a sequence of generalized random coding schemes $\{({\cal L}^{(N)}, \gamma^{(N)} )\}$ following an asymptotic input distribution $P_{X|r}$. Assume that $P_{X|r}$ is continuous in $r$ except at a finite number of points. The following standard communication rate region $R$ is asymptotically achievable,
\begin{equation}
R=\left\{ r| \mbox{such that, either } r=0 \mbox{ or } r<I_r(X; Y)\right\},
\end{equation}
where the mutual information $I_r(X; Y)$ is computed using input distribution $P_{X|r}$.

Furthermore, for any sequence of random coding schemes following asymptotic input distribution $P_{X|r}$, assume that $\tilde{R}$ is an asymptotically achievable rate region. Let $\tilde{r}$ be an arbitrary rate inside $\tilde{R}$ in the sense that we can find $\delta>0$ with $r\in \tilde{R}$ for all $\tilde{r}\le r \le \tilde{r}+\delta$. If the asymptotic conditional input distribution $P_{X|r}$ is continuous in $r$ at $\tilde{r}$, then we must have
\begin{equation}
\tilde{r} \le I_{\tilde{r}}(X; Y).
\end{equation}
$\QED$
\end{theorem}

Theorem \ref{Theorem1} is implied by Theorem \ref{Theorem2} given in Section \ref{SectionIII}.

Because the receiver has the option of reporting a collision even if it can decode the message reliably, any region contained inside an asymptotically achievable rate region is also asymptotically achievable. To understand the practical meaning of the achievable rate region result, assume that practical communication rates are defined as a function of $W$ at the transmitter. For any subset of communication rates, if the standard rates of all the corresponding messages are asymptotically contained in the achievable rate region, then the subset of communication rates are asymptotically achievable in the same sense as specified in Section \ref{SectionII.A}.

It is important to note that the achievable rate region characterized in Theorem \ref{Theorem1} is significantly different from a Shannon information rate region \cite{ref Shannon48}. For reliable communication in the classical Shannon sense, information rate equals the normalized log of the codebook size. Given the codeword length, the rate must be predetermined since both the transmitter and the receiver need to know the codebook before message transmission is started\footnote{In a rateless communication model, the total number of data bits encoded at the transmitter, which corresponds to the log of the codebook size, must be predetermined.}. Collision or erasure detection is therefore not needed if the receiver knows the channel \cite{ref Cover05}. In our coding approach for random access communication, the codebook contains a large number of codewords. The codewords are categorized into subsets with different rate values. If the transmitter does not know the channel and chooses its rate arbitrarily, it becomes the receiver's responsibility to {\it detect} whether the transmitted codeword is inside/outside the achievable rate region. This is a fundamental functionality required in random access communication but not seen in classical Information Theoretic communication.

By regarding collision report at the receiver as message erasure, the new coding approach shares certain similarity with error and erasure decoding \cite{ref Forney68} in a classical communication system. However, there is also a fundamental difference. In classical communication, if the codebook is optimized with information rate below the channel capacity, then both the error and the erasure probabilities diminish asymptotically for {\it all} codewords in the codebook. In our coding approach, however, the normalized log of the codebook size can be significantly higher than the channel capacity. In this case, codewords are essentially partitioned into two exclusive subsets according to their associated rate values. For codewords whose rates are inside the achievable rate region, the probability of successful decoding approaches one asymptotically. For codewords whose rates are outside the region, however, the probability of message erasure goes to one asymptotically. Such codeword categorization and codeword-dependent asymptotic decoding behavior are not seen in classical communication systems.

Note that, in Theorem \ref{Theorem1}, the asymptotic input distribution $P_{X|r}$ could be discontinuous in $r$. For random access communication, the inclusion of a possible discontinuous input distribution $P_{X|r}$ is necessary. Consider the last example of a coding scheme given in Section \ref{SectionII.A}. Each codebook in the library is partitioned into three codeword classes. The codeword classes contain $1$, $2^{NR_0}$, $2^{NR_1}$ codewords, respectively. We have explained that, practically, these codeword classes correspond to the options of encoding $0$, $NR_0$, and $NR_1$ data bits into a packet. Suppose that the communication channel is Gaussian. For the three encoding options, a common choice of the transmitter could be to set the input distributions as Gaussian corresponding to three different transmission powers, $P=0$, $P=P_0$, and $P=P_1$. Let us denote the three Gaussian input distributions by ${\cal N}(0, 0)$, ${\cal N}(0, P_0)$, ${\cal N}(0, P_1)$. If we characterize the asymptotic input distribution as a function of the {\it standard communication rate}, we obtain
\begin{equation}
P_{X|r}=\left\{\begin{array}{ll} {\cal N}(0, 0) & r=0 \\ {\cal N}(0, P_0) & 0<r\le R_0 \\ {\cal N}(0, P_1) & R_0<r \le R_1 \end{array} \right.
\end{equation}
In this example, $P_{X|r}$ is discontinuous in $r$ at $r=0$ and $r=R_0$.

\section{A New Packet Coding Approach -- Random Multiple Access Communication}
\label{SectionIII}

In this section, we extend the previously developed coding theorem to a $K$-user, symbol synchronous, random multiple access system over a discrete-time memoryless channel. The channel is modeled by a conditional distribution function $P_{Y|X_1, \dots, X_K}$, where $X_i\in {\cal X}_i$ is the channel input symbol of user $i$ with ${\cal X}_i$ being the input alphabet, and $Y\in {\cal Y}$ is the channel output symbol with ${\cal Y}$ being the output alphabet. To simplify the discussion, we assume that ${\cal Y}$ and ${\cal X}_i$, for all $i$, are finite. Extending the results to continuous channels is straightforward. As in Section \ref{SectionII}, we assume that time is slotted with each slot being equal to $N$ symbol durations, which is also the length of a packet.

We assume that each user, say user $i$, is equipped with a generalized random coding scheme $({\cal L}_i, \gamma_i )$. Each codebook in the library contains $2^{NR_{\max}}$ codewords of length $N$, where $R_{\max}$ is a pre-determined large constant whose particular value is not important. At the beginning of each time slot, user $i$ randomly generates a codebook index $\theta_i$, and uses the corresponding codebook in its library to encode a macro message $W_i$ into a codeword. We assume that the channel is known both at the transmitters and at the receiver. The receiver also knows the particular codebook chosen by each user, and this can be achieved by sharing the random codebook generation algorithms with the receiver. However, we assume that communication rates of the users are shared neither among each other nor with the receiver.

We use bold-font characters to denote vectors whose $i$th elements are the corresponding variables of user $i$. For example, $\mbox{\boldmath  ${\cal L}$}$ represents the vector of code libraries of the users. Also, $\mbox{\boldmath  $\theta$}$ denotes the random index vector, $\mbox{\boldmath  ${\cal C}$}_{\mbox{\scriptsize \boldmath  $\theta$}}$ denotes the codebook vector, $\mbox{\boldmath  $W$}$ denotes the message vector, $\mbox{\boldmath  $r$}(\mbox{\boldmath  $W$})$ denotes the standard rate vector, and $\mbox{\boldmath  $P$}_{\mbox{\scriptsize \boldmath  $X$}|\mbox{\scriptsize \boldmath  $r$}}$ denotes the asymptotic input distributions of the random coding schemes, etc.

\subsection{Collision Detection for All Users}
\label{SectionIII.A}

Let $\mbox{\boldmath  $W$}$ be the transmitted message vector, encoded using codebook $\mbox{\boldmath  ${\cal C}$}_{\mbox{\scriptsize \boldmath  $\theta$}}$. Assume that, upon observing the received channel symbols, the receiver either outputs message estimate $\hat{\mbox{\boldmath  $W$}}$ for all users, or reports a collision. We define $P_{e|\mbox{\scriptsize \boldmath  $\theta$}}(\mbox{\boldmath  ${\cal C}$}_{\mbox{\scriptsize \boldmath  $\theta$}}, \mbox{\boldmath  $W$})=Pr\{\hat{\mbox{\boldmath  $W$}} \ne \mbox{\boldmath  $W$} \}$ as the probability, conditioned on $\mbox{\boldmath  $\theta$}$, that the receiver is not able to recover the message vector $\mbox{\boldmath  $W$}$. Define $P_{c|\mbox{\scriptsize \boldmath  $\theta$}}(\mbox{\boldmath  ${\cal C}$}_{\mbox{\scriptsize \boldmath  $\theta$}}, \mbox{\boldmath  $W$})=Pr\{\mbox{collision}\}$ as the conditional probability that the receiver reports a collision. Assume random coding schemes $(\mbox{\boldmath  ${\cal L}$}, \mbox{\boldmath  $\gamma$})$. Let $\mbox{\boldmath  $\theta$}$ be drawn independently according to $\mbox{\boldmath  $\gamma$}$. We define $P_e(\mbox{\boldmath  $W$})$ and $P_c(\mbox{\boldmath  $W$})$ as the unconditional error probability and the unconditional collision probability of message $\mbox{\boldmath  $W$}$, respectively. That is,
\begin{equation}
P_e(\mbox{\boldmath  $W$})=E_{\mbox{\scriptsize \boldmath  $\theta$}}[P_{e|\mbox{\scriptsize \boldmath  $\theta$}}(\mbox{\boldmath  ${\cal C}$}_{\mbox{\scriptsize \boldmath  $\theta$}}, \mbox{\boldmath  $W$} )], \mbox{  } P_c(\mbox{\boldmath  $W$})=E_{\mbox{\scriptsize \boldmath  $\theta$}}[P_{c|\mbox{\scriptsize \boldmath  $\theta$}}(\mbox{\boldmath  ${\cal C}$}_{\mbox{\scriptsize \boldmath  $\theta$}}, \mbox{\boldmath  $W$})].
\end{equation}

\begin{definition}{\label{Definition5}}
Consider random multiple access communication using a sequence of random coding schemes $\{(\mbox{\boldmath  ${\cal L}$}^{(N)}, \mbox{\boldmath  $\gamma$}^{(N)})\}$, where $(\mbox{\boldmath  ${\cal L}$}^{(N)}, \mbox{\boldmath  $\gamma$}^{(N)})$ is a vector of generalized random coding schemes with each codebook in  $\mbox{\boldmath  ${\cal L}$}^{(N)}$ containing $2^{NR_{\max}}$ codewords of length $N$. Let $\mbox{\boldmath  $R$}$ be a region of standard rate vectors. Let $\mbox{\boldmath  $R$}_c$ be the closure of $\mbox{\boldmath  $R$}$. We say $\mbox{\boldmath  $R$}$ is asymptotically achievable if there exists a sequence of decoding algorithms under which the following two conditions are satisfied. First, for all message sequences $\{\mbox{\boldmath  $W$}^{(N)}\}$ with $\mbox{\boldmath  $r$}(\mbox{\boldmath  $W$}^{(N)})\in \mbox{\boldmath  $R$}$ for all $N$ and $\lim_{N\to\infty}\mbox{\boldmath  $r$}(\mbox{\boldmath  $W$}^{(N)})\in \mbox{\boldmath  $R$}$, we have $\lim_{N\to \infty} P_e(\mbox{\boldmath  $W$}^{(N)})=0$. Second, for all message sequences $\{\mbox{\boldmath  $W$}^{(N)}\}$ with $\mbox{\boldmath  $r$}(\mbox{\boldmath  $W$}^{(N)})\not\in \mbox{\boldmath  $R$}_c$ for all $N$ and $\lim_{N\to\infty}\mbox{\boldmath  $r$}(\mbox{\boldmath  $W$}^{(N)})\not\in \mbox{\boldmath  $R$}_c$, we have $\lim_{N\to \infty} P_c(\mbox{\boldmath  $W$}^{(N)})=1$. $\QED$
\end{definition}

The following theorem characterizes the asymptotically achievable rate region of a random multiple access system.

\begin{theorem}{\label{Theorem2}}
Consider random multiple access communication over a discrete-time memoryless channel $P_{Y|X_1, \dots, X_K}$ using a sequence of random coding schemes $\{(\mbox{\boldmath  ${\cal L}$}^{(N)}, \mbox{\boldmath  $\gamma$}^{(N)})\}$. Assume that $\{(\mbox{\boldmath  ${\cal L}$}^{(N)}, \mbox{\boldmath  $\gamma$}^{(N)})\}$ follows asymptotic distribution $\mbox{\boldmath  $P$}_{\mbox{\scriptsize \boldmath  $X$}|\mbox{\scriptsize \boldmath  $r$}}$. Assume that, for any user $i$, $P_{X_i|r_i}$ is only discontinuous in $r_i$ at a finite number of points. The following standard communication rate region $\mbox{\boldmath  $R$}$ is asymptotically achievable.
\begin{equation}
\mbox{\boldmath  $R$}=\left\{\mbox{\boldmath  $r$}\left| \forall S \subseteq \{1, \dots, K\}, \mbox{ either } r_{i\in S} =0, \mbox{ or }
\sum_{i\in S} r_i < I_{\mbox{\scriptsize \boldmath  $r$}} ( \mbox{\boldmath  $X$}_{i\in S} ; Y |\mbox{\boldmath  $X$}_{i \not\in S}) \right.\right\},
\label{MutualInformationInequality}
\end{equation}
where the mutual information $I_{\mbox{\scriptsize \boldmath  $r$}}( \mbox{\boldmath  $X$}_{i\in S} ; Y |\mbox{\boldmath  $X$}_{i \not\in S})$ is computed using input distribution $\mbox{\boldmath  $P$}_{\mbox{\scriptsize \boldmath  $X$}|\mbox{\scriptsize \boldmath  $r$}}$.

Furthermore, for any sequence of random coding schemes following the asymptotic conditional input distribution $\mbox{\boldmath  $P$}_{\mbox{\scriptsize \boldmath  $X$}|\mbox{\scriptsize \boldmath  $r$}}$, assume that $\tilde{\mbox{\boldmath  $R$}}$ is an asymptotically achievable rate region. Let $\tilde{\mbox{\boldmath  $r$}}$ be an arbitrary rate vector inside $\tilde{\mbox{\boldmath  $R$}}$, in the sense that we can find $\delta>0$ with $\mbox{\boldmath  $r$} \in \tilde{\mbox{\boldmath  $R$}}$ for all $\tilde{\mbox{\boldmath  $r$}}\le \mbox{\boldmath  $r$} \le \tilde{\mbox{\boldmath  $r$}}+\delta \mbox{\boldmath  $1$}$, where $\mbox{\boldmath  $1$}$ is a vector of all ones. Let $S \subseteq \{1, \dots, K\}$ be an arbitrary user subset. If the asymptotic conditional input distribution $\mbox{\boldmath  $P$}_{\mbox{\scriptsize \boldmath  $X$}|\mbox{\scriptsize \boldmath  $r$}}$ is continuous in $\mbox{\boldmath  $r$}_{i\in S}$ at $\tilde{\mbox{\boldmath  $r$}}$, then we must have
\begin{equation}
\sum_{i\in S} \tilde{r}_i \le I_{\tilde{\mbox{\scriptsize \boldmath  $r$}}} ( \mbox{\boldmath  $X$}_{i\in S} ; Y |\mbox{\boldmath  $X$}_{i \not\in S}).
\label{MutualInformationInequalityConverse}
\end{equation}
$\QED$
\end{theorem}

The achievability part of Theorem \ref{Theorem2} is implied by Theorem \ref{Theorem3} which is provided later in this section. The proof of Theorem \ref{Theorem3} is given in Appendix \ref{ProofTheorem3} and the converse part of Theorem \ref{Theorem2} is proved in Appendix \ref{ProofTheorem2}.

When the asymptotic conditional input distribution $\mbox{\boldmath  $P$}_{\mbox{\scriptsize \boldmath  $X$}|\mbox{\scriptsize \boldmath  $r$}}$ is not a function of $\mbox{\boldmath  $r$}$, i.e. codewords of each user are generated according to the same input distribution, the achievable rate region $\mbox{\boldmath  $R$}$ given in (\ref{MutualInformationInequality}) becomes
\begin{equation}
\mbox{\boldmath  $R$}=\left\{\mbox{\boldmath  $r$}\left| \forall S \subseteq \{1, \dots, K\}, \sum_{i\in S} r_i < I( \mbox{\boldmath  $X$}_{i\in S} ; Y |\mbox{\boldmath  $X$}_{i \not\in S}) \right.\right\}.
\label{MutualInformationInequality0}
\end{equation}
This is identical to the Shannon information rate region without a convex hull operation for the multiple access channel under a given input distribution \cite{ref Cover05}.

Note that in a classical channel coding scheme, users jointly determine their information rates by choosing codebooks with appropriate numbers of codewords. To decode the messages, the receiver essentially needs to search all codewords (e.g. maximum likelihood decoding or typical sequence decoding) throughout the codebook \cite{ref Cover05}. Random access communication does not assume joint rate determination among users \cite{ref Luo06}. Therefore the users cannot know a priori if their rate vector is within the achievable rate region or not. In this sense, the situation is similar to the single-user case discussed in Section \ref{SectionII}. In our coding approach, the codebooks of the users contain large numbers of codewords. However, the codewords are indexed by their standard rate parameters and the receiver only searches for appropriate codewords within the achievable rate region, which can be regarded as a subset of the joint codewords in the codebooks. The receiver reports a collision if an appropriate codeword cannot be found.

\subsection{Collision Detection for Each Individual User}
\label{SectionIII.B}

In Section \ref{SectionIII.A}, we assumed that the receiver should either output reliable message estimates for {\it all} users, or report a collision. When multiple users compete for the channel, it is often the case that the receiver may only be interested in recovering messages from a subset of users. Correspondingly, a collision should only be reported when messages from the users of interest are not decodable. Let us first assume that the receiver only wants to recover the message from a particular user, say user $k$. Given that message $\mbox{\boldmath  $W$}$ is transmitted over the multiple access channel using codebook $\mbox{\boldmath  ${\cal C}$}_{\mbox{\scriptsize \boldmath  $\theta$}}$, the receiver either outputs a message estimate $\hat{W}_k$ for user $k$, or reports a collision.

Before we proceed further, it is important to note that, even though the receiver is only interested in decoding the message of user $k$, whether reliable message recovery is possible for user $k$ still can depend on the {\it transmitted} messages of {\it all} users. Therefore, when we define the error probabilities and the achievable rate region, the rates of {\it all} users are still involved.

Define $P_{ek|\mbox{\scriptsize \boldmath  $\theta$}}(\mbox{\boldmath  ${\cal C}$}_{\mbox{\scriptsize \boldmath  $\theta$}}, \mbox{\boldmath  $W$} )=Pr\{\hat{W}_k\ne W_k\}$ as the probability, conditioned on $\mbox{\boldmath  $\theta$}$ that the receiver is not able to recover user $k$'s message $W_k$. Define $P_{ck|\mbox{\scriptsize \boldmath  $\theta$}}(\mbox{\boldmath  ${\cal C}$}_{\mbox{\scriptsize \boldmath  $\theta$}}, \mbox{\boldmath  $W$} )=Pr\{\mbox{collision}\}$ as the conditional probability that the receiver reports a collision for user $k$. We define $P_{ek}(\mbox{\boldmath  $W$})$ and $P_{ck}(\mbox{\boldmath  $W$})$ as the unconditional error probability and the unconditional collision probability of message $\mbox{\boldmath  $W$}$ for user $k$; that is
\begin{equation}
P_{ek}(\mbox{\boldmath  $W$})=E_{\mbox{\scriptsize \boldmath  $\theta$}}[P_{ek|\mbox{\scriptsize \boldmath  $\theta$}}(\mbox{\boldmath  ${\cal C}$}_{\mbox{\scriptsize \boldmath  $\theta$}}, \mbox{\boldmath  $W$} )], \mbox{  } P_{ck}(\mbox{\boldmath  $W$})=E_{\mbox{\scriptsize \boldmath  $\theta$}}[P_{ck|\mbox{\scriptsize \boldmath  $\theta$}}(\mbox{\boldmath  ${\cal C}$}_{\mbox{\scriptsize \boldmath  $\theta$}}, \mbox{\boldmath  $W$} )].
\label{PekPck}
\end{equation}

\begin{definition}{\label{Definition6}}
Consider random multiple access communication using a sequence of random coding schemes $\{(\mbox{\boldmath  ${\cal L}$}^{(N)}, \mbox{\boldmath  $\gamma$}^{(N)})\}$, where $(\mbox{\boldmath  ${\cal L}$}^{(N)}, \mbox{\boldmath  $\gamma$}^{(N)})$ is a generalized random coding scheme with each codebook in $\mbox{\boldmath  ${\cal L}$}^{(N)}$ containing $2^{NR_{\max}}$ codewords of length $N$. Let $\mbox{\boldmath  $R$}_k$ be a region of standard rate vectors. Let $\mbox{\boldmath  $R$}_{kc}$ be the closure of $\mbox{\boldmath  $R$}_k$. We say $\mbox{\boldmath  $R$}_k$ is asymptotically achievable for user $k$ if there exists a sequence of decoding algorithms under which the following two conditions are satisfied. First, for all message sequences $\{\mbox{\boldmath  $W$}^{(N)}\}$ with $\mbox{\boldmath  $r$}(\mbox{\boldmath  $W$}^{(N)})\in \mbox{\boldmath  $R$}_k$ for all $N$ and $\lim_{N\to\infty}\mbox{\boldmath  $r$}(\mbox{\boldmath  $W$}^{(N)})\in \mbox{\boldmath  $R$}_k$, we have $\lim_{N\to \infty} P_{ek}(\mbox{\boldmath  $W$}^{(N)})=0$. Second, for all message sequences $\{\mbox{\boldmath  $W$}^{(N)}\}$ with $\mbox{\boldmath  $r$}(\mbox{\boldmath  $W$}^{(N)})\not\in \mbox{\boldmath  $R$}_{kc}$ for all $N$ and $\lim_{N\to\infty}\mbox{\boldmath  $r$}(\mbox{\boldmath  $W$}^{(N)})\not\in \mbox{\boldmath  $R$}_{kc}$, we have $\lim_{N\to \infty} P_{ck}(\mbox{\boldmath  $W$}^{(N)})=1$. $\QED$
\end{definition}

The following theorem gives an achievable rate region for user $k$.

\begin{theorem}{\label{Theorem3}}
Consider random multiple access communication over a discrete-time memoryless multiple access channel $P_{Y|X_1, \dots, X_K}$ using a sequence of random coding schemes $\{(\mbox{\boldmath  ${\cal L}$}^{(N)}, \mbox{\boldmath  $\gamma$}^{(N)})\}$. Assume that $\{(\mbox{\boldmath  ${\cal L}$}^{(N)}, \mbox{\boldmath  $\gamma$}^{(N)})\}$ follows asymptotic input distributions $\mbox{\boldmath  $P$}_{\mbox{\scriptsize \boldmath  $X$}|\mbox{\scriptsize \boldmath  $r$}}$. For any user $i$, $P_{X_i|r_i}$ is only discontinuous in $r_i$ at a finite number of points. The following standard rate region $\mbox{\boldmath  $R$}_k$ is asymptotically achievable for user $k$, namely
\begin{equation}
\mbox{\boldmath  $R$}_{k}=\left\{\mbox{\boldmath  $r$}\left| \begin{array}{l} \mbox{such that } \forall S \subseteq \{1, \dots, K\}, k\in S, \mbox{ either } r_k=0, \\ \mbox{or } \exists \tilde{S}\subseteq S, k\in \tilde{S}, \mbox{ such that}, \sum_{i\in \tilde{S}}r_i < I_{\mbox{\scriptsize \boldmath  $r$}} ( \mbox{\boldmath  $X$}_{i \in \tilde{S}} ; Y |\mbox{\boldmath  $X$}_{i \not\in S})  \end{array} \right.\right\},
\label{MutualInformationInequality1}
\end{equation}
where the mutual information $I_{\mbox{\scriptsize \boldmath  $r$}}( \mbox{\boldmath  $X$}_{i\in \tilde{S}} ; Y |\mbox{\boldmath  $X$}_{i \not\in S})$ is computed using input distribution $\mbox{\boldmath  $P$}_{\mbox{\scriptsize \boldmath  $X$}|\mbox{\scriptsize \boldmath  $r$}}$.
$\QED$
\end{theorem}

The proof of Theorem \ref{Theorem3} is given in Appendix \ref{ProofTheorem3}.

Theorem \ref{Theorem3} can be easily extended to the case when the receiver is interested in recovering messages from a subset of users, as shown in the next theorem.

\begin{theorem}{\label{Theorem4}}
Consider random multiple access communication over a discrete-time memoryless multiple access channel $P_{Y|X_1, \dots, X_K}$ using a sequence of random coding schemes $\{(\mbox{\boldmath  ${\cal L}$}^{(N)}, \mbox{\boldmath  $\gamma$}^{(N)})\}$. Assume that $\{(\mbox{\boldmath  ${\cal L}$}^{(N)}, \mbox{\boldmath  $\gamma$}^{(N)})\}$ follows the asymptotic input distributions $\mbox{\boldmath  $P$}_{\mbox{\scriptsize \boldmath  $X$}|\mbox{\scriptsize \boldmath  $r$}}$. Assume that, for any user $i$, the quantity $P_{X_i|r_i}$ is only discontinuous in $r_i$ at a finite number of points. Let $S_0 \subseteq \{1, \dots, K\}$ be a user subset. The following rate region $\mbox{\boldmath  $R$}_{S_0}$ is asymptotically achievable for users $k\in S_0$.
\begin{equation}
\mbox{\boldmath  $R$}_{S_0}=\left\{\mbox{\boldmath  $r$}\left| \begin{array}{l}  \mbox{such that } \forall S \subseteq \{1, \dots, K\}, S \cap S_0 \ne \phi, \exists  \tilde{S}, S \cap S_0 \subseteq \tilde{S} \subseteq S, \\ \mbox{such that, either } r_{i\in \tilde{S}}=0,  \mbox{ or } \sum_{i\in \tilde{S}}r_i < I_{\mbox{\scriptsize \boldmath  $r$}} ( \mbox{\boldmath  $X$}_{i \in \tilde{S}} ; Y |\mbox{\boldmath  $X$}_{i \not\in S})  \end{array} \right.\right\},
\label{MutualInformationInequality2}
\end{equation}
where the mutual information $I_{\mbox{\scriptsize \boldmath  $r$}}( \mbox{\boldmath  $X$}_{j\in \tilde{S}} ; Y |\mbox{\boldmath  $X$}_{k \not\in S})$ is computed using the input distribution $\mbox{\boldmath  $P$}_{\mbox{\scriptsize \boldmath  $X$}|\mbox{\scriptsize \boldmath  $r$}}$.
$\QED$
\end{theorem}

The proof of Theorem \ref{Theorem4} is given in Appendix \ref{ProofTheorem4}.

Note that when $S_0=\{1, \dots, K\}$, the region $\mbox{\boldmath  $R$}_{S_0}$ given in (\ref{MutualInformationInequality2}) is equal to the rate region $\mbox{\boldmath  $R$}$ given in (\ref{MutualInformationInequality0}).

\section{Simple Examples}
\label{SectionIV}

In this section, we illustrate the achievable rate region results in two simple example systems.

{\bf Example 1: } $\quad$ Consider a $K$-user random multiple access system over a memoryless Gaussian channel modeled by
\begin{equation}
Y=\sum_{i=1}^{K}X_i + V.
\end{equation}
where $V$ is a white Gaussian noise with zero mean and variance $N_0$.

Assume that the input distribution of user $k$ is Gaussian with zero mean and variance $P_k$, irrespective of the rate parameter. Assume that the receiver wants to recover messages of all users. According to Theorem \ref{Theorem1}, the following rate region is asymptotically achievable
\begin{equation}
\mbox{\boldmath  $R$}=\left\{\mbox{\boldmath  $r$}\left| \forall S \subseteq \{1, \dots, K\}, \sum_{i\in S}r_i < \frac{1}{2}\log \left(1+ \frac{\sum_{i\in S}P_i}{N_0}\right) \right.\right\}.
\label{GaussianRateRegion}
\end{equation}
Note that the achievable rate region $\mbox{\boldmath  $R$}$ is identical to the Shannon channel capacity region \cite{ref Cover05}.

If for each $k$, the input distribution is Gaussian with zero mean and variance $P_k$ for any non-zero rate, and user $k$ idles at rate zero, then the achievable rate region is still given by (\ref{GaussianRateRegion}).

{\bf Example 2: } $\quad$ Consider a $K$-user random multiple access system over a memoryless symbol collision channel. We define an $n$th order symbol collision channel as follows. The channel input alphabet of any user is given by ${\cal X}=\{0, 1, \dots, 2^n \}$, where $0$ represents an idle symbol. The channel output alphabet is given by ${\cal Y}=\{0, 1, \dots, 2^n, c\}$, where $c$ represents a collision symbol. If all users idle, the receiver receives an idle symbol, $Y=0$; if only one user, say user $k$, transmits a non-zero symbol $X_k$, the receiver receives $Y=X_k$; if multiple users transmit non-zero symbols, the receiver receives $Y=c$, i.e., a collision symbol. We assume that in all input distributions the non-zero symbols always have equal probabilities. Consequently, an input distribution $P_{X_k|r_k}$ can be characterized through a single parameter $p(r_k)$, which is the probability that any particular symbol in the transmitted codeword takes a non-zero value.

\begin{proposition}{\label{Proposition1}}
Assume that the conditional input distribution of user $k$, for all $k$, is given by
\begin{equation}
P_{X_k|r_k}=\left\{\begin{array}{cl}1-\sqrt{r_k/n} & \mbox{for } X_k=0 \\ \frac{1}{2^n}\sqrt{r_k/n} & \mbox{for } X_k=j \in \{1, \dots, 2^n\} \end{array} \right. .
\end{equation}
In other words, let $p(r_k)=\sqrt{r_k/n}$. Assume that the receiver wants to recover the messages of all users. The following rate region $\mbox{\boldmath  $R$}$ is asymptotically achievable.
\begin{equation}
\mbox{\boldmath  $R$}=\left\{\mbox{\boldmath  $r$}\left| \sum_{i=1}^{K}\sqrt{r_k/n} < 1 \right.\right\}.
\label{RateRegion1}
\end{equation}
$\QED$
\end{proposition}

Proposition \ref{Proposition1} is proven in Appendix \ref{ProofProposition1}.

Note that when $K=2$, the rate region $\mbox{\boldmath  $R$}$ given in (\ref{RateRegion1}) equals the random multiple access throughput and stability regions of the collision channel \cite{ref Rao88}, which also approaches the asynchronous information capacity region of the collision channel as $n\to \infty$ \cite{ref Hui84}. Proposition \ref{Proposition1} therefore motivates the question whether there is a fundamental connection between the achievable rate region studied in this paper and the throughput, stability, information capacity regions of the random multiple access channel discussed in \cite{ref Luo06}. Obtaining a theoretical answer to this question is an important issue for future research.

In both examples, the achievable rate region $\mbox{\boldmath  $R$}$ is coordinate convex in the sense that if $\hat{\mbox{\boldmath  $r$}} \in \mbox{\boldmath  $R$}$ then $\mbox{\boldmath  $r$}\in \mbox{\boldmath  $R$}$ for all $\mbox{\boldmath  $r$}\le \hat{\mbox{\boldmath  $r$}} $. However, in general, depending on the choice of the asymptotic input distribution, the region may not be continuous in the sense that not every point pair in the region is connected by a continuous path in the region. In this case, the region is certainly not coordinate convex.

\section{Further Discussions}
\label{SectionV}
Throughout our analysis, we obtained asymptotic results by taking the packet length to infinity. It is unfortunately the nature of random access systems that their packets are often substantially short. Therefore, it is necessary to investigate the achievable rate and error performance under the assumption of a finite packet length. This is similar to the channel dispersion analysis presented in \cite{ref Polyanskiy10} for classical communication systems. Alternatively, one can also characterize the achievable rate and error performance tradeoff in a way similarly to the error exponent analysis for classical channel coding \cite{ref Gallager65}. An example of such derivation can be found in \cite{ref Wang10}.

We have assumed that the multiple access channel is fully known at the receiver. Even though users do not pre-share their rate information, their codebooks are assumed known at the receiver. To extend our results to other random access scenarios, one has to carefully examine whether these assumptions are still valid or reasonable. If not, deriving coding performance under various channel and codebook information availability assumptions becomes the key challenge.

The coding approach studied in this paper enabled a network user to choose its communication rate by encoding a variable number of data bits into a packet. How should a user determine its communication rate in a particular time slot, however, is a challenging MAC layer problem. Nevertheless, compared to the packet-based channel model used in classical MAC layer performance analysis \cite{ref Luo06}, performance limitations of the introduced coding approach provide a useful bridge to support the joint optimization of MAC layer networking and physical layer channel coding.

\section{Conclusion}
\label{SectionVI}

We proposed a new channel coding approach for coding within each packet in a random access system with bursty traffic. The coding approach enabled each user to choose its communication rate without pre-sharing the rate information with the receiver and other users. Performance of the coding approach is characterized by an achievable region defined on the communication rates. The receiver is required to output reliable message estimates if the communication rates are inside the achievable rate region; the receiver should report a collision if the rates are outside the region. We showed that the maximum achievable rate region of random coding schemes takes a form similar to the Shannon information rate region without a convex hull operation. The achievable rate region when the receiver is interested in recovering messages only from a subset of users is also obtained.

There are numerous questions left open that would further tighten the connection between random access networking and Information Theory. We believe that our approach contributes an important component to that connection by distinguishing the issues of reliable communication and reliable collision detection in a rigorous manner.

\appendix
\subsection{Proof of Theorem \ref{Theorem2}}
\label{ProofTheorem2}
\begin{proof}
Note that the achievability part of Theorem \ref{Theorem2} is implied by Theorem \ref{Theorem3}. Here we prove the converse part of Theorem \ref{Theorem2}.

Let $\tilde{\mbox{\boldmath  $R$}}$ be an asymptotically achievable rate region. Let $\tilde{\mbox{\boldmath  $r$}} \in \tilde{\mbox{\boldmath  $R$}}$ be a rate vector inside $\tilde{\mbox{\boldmath  $R$}}$. We can find a $\delta >0$ such that $\mbox{\boldmath  $r$}\in \tilde{\mbox{\boldmath  $R$}}$ for all $\tilde{\mbox{\boldmath  $r$}} \le \mbox{\boldmath  $r$} \le \tilde{\mbox{\boldmath  $r$}}+\delta \mbox{\boldmath  $1$}$.

Let $S\subseteq \{1, \dots, K\}$ be a given user subset. If the asymptotic distribution $\mbox{\boldmath  $P$}_{\mbox{\scriptsize \boldmath  $X$}|\mbox{\scriptsize \boldmath  $r$}}$, and hence the entropy functions, are continuous in $\mbox{\boldmath  $r$}_{i\in S}$ at $\tilde{\mbox{\boldmath  $r$}}$, we can find a small enough $\delta>0$ and a bound $u_{\delta}>0$ with $\lim_{\delta\to 0}u_{\delta}=0$, such that the following inequality holds for all rates $\mbox{\boldmath  $r$}$, with $\mbox{\boldmath  $r$}_{i\not\in S}=\tilde{\mbox{\boldmath  $r$}}_{i\not\in S}$ and $\tilde{\mbox{\boldmath  $r$}}_{i\in S} \le \mbox{\boldmath  $r$}_{i\in S} < \tilde{\mbox{\boldmath  $r$}}_{i\in S}+\delta \mbox{\boldmath  $1$}_{i\in S}$:
\begin{equation}
|I_{\mbox{\scriptsize \boldmath  $r$}}(\mbox{\boldmath  $X$}_{i\in S}; Y |\mbox{\boldmath  $X$}_{i\not\in S} ) -I_{\tilde{\mbox{\scriptsize \boldmath  $r$}}}(\mbox{\boldmath  $X$}_{i\in S}; Y |\mbox{\boldmath  $X$}_{i\not\in S} ) | \le u_{\delta},
\end{equation}
where the mutual information $I_{\mbox{\scriptsize \boldmath  $r$}}(\mbox{\boldmath  $X$}_{i\in S}; Y |\mbox{\boldmath  $X$}_{i\not\in S} )$ is evaluated using input distribution $\mbox{\boldmath  $P$}_{\mbox{\scriptsize \boldmath  $X$}|\mbox{\scriptsize \boldmath  $r$}}$ and the mutual information $I_{\tilde{\mbox{\scriptsize \boldmath  $r$}}}(\mbox{\boldmath  $X$}_{i\in S}; Y |\mbox{\boldmath  $X$}_{i\not\in S} )$ is evaluated using input distribution $\mbox{\boldmath  $P$}_{\mbox{\scriptsize \boldmath  $X$}|\tilde{\mbox{\scriptsize \boldmath  $r$}}}$.

Let $\mbox{\boldmath  $W$}^{(N)}$ be the actual source message. We assume that $\mbox{\boldmath  $W$}^{(N)}$ is generated randomly according to a uniform distribution under the condition that $\mbox{\boldmath  $r$}_{i\not\in S}(\mbox{\boldmath  $W$}^{(N)})=\tilde{\mbox{\boldmath  $r$}}_{i\not\in S}$ and $\tilde{\mbox{\boldmath  $r$}}_{i\in S} \le \mbox{\boldmath  $r$}_{i\in S}(\mbox{\boldmath  $W$}^{(N)}) < \tilde{\mbox{\boldmath  $r$}}_{i\in S} + \delta \mbox{\boldmath  $1$}_{i\in S}$. Assume that the codewords $\mbox{\boldmath  $X$}_{i\not\in S}^{(N)}$ are known to the receiver. Let $\hat{\mbox{\boldmath  $W$}}^{(N)}$ be the message estimate generated at the receiver. Define $P_e^{(N)}=Pr\{\hat{\mbox{\boldmath  $W$}}^{(N)} \ne \mbox{\boldmath  $W$}^{(N)} \}$ as the error probability. Note that
\begin{eqnarray}
\lim_{N\to \infty} \frac{1}{N}H(\mbox{\boldmath  $W$}^{(N)}_{i\in S})&&=\lim_{N\to \infty} \frac{1}{N} \sum_{i\in S} \log_2\left( 2^{N(\tilde{r}_i+\delta)}  - 2^{N\tilde{r}_i} \right)      = \sum_{i\in S} (\tilde{r}_i+\delta).
\end{eqnarray}
We assume that $N$ is large enough such that $H(\mbox{\boldmath  $W$}^{(N)}_{i\in S})\ge N \sum_{i \in S} \tilde{r}_i$. According to Fano's inequality \cite{ref Fano61}, we have
\begin{eqnarray}
&& \sum_{i \in S} \tilde{r}_i \le \frac{1}{N} H(\mbox{\boldmath  $W$}^{(N)}_{i\in S})       = \frac{1}{N} H(\mbox{\boldmath  $W$}^{(N)}_{i\in S}|\mbox{\boldmath  $W$}^{(N)}_{i\not\in S}) \nonumber \\
&& \quad =\frac{1}{N} H(\mbox{\boldmath  $W$}^{(N)}_{i\in S}|\mbox{\boldmath  $W$}^{(N)}_{i\not\in S}, \hat{\mbox{\boldmath  $W$}}^{(N)}_{i\in S})+ \frac{1}{N}I(\mbox{\boldmath  $W$}^{(N)}_{i\in S} ; \hat{\mbox{\boldmath  $W$}}^{(N)}_{i\in S}|\mbox{\boldmath  $W$}^{(N)}_{i\not\in S}) \nonumber \\
&& \quad < \frac{1}{N} + \frac{1}{N}P_e^{(N)} H(\mbox{\boldmath  $W$}^{(N)}_{i\in S})  + \frac{1}{N}I(\mbox{\boldmath  $W$}^{(N)}_{i\in S} ; \hat{\mbox{\boldmath  $W$}}^{(N)}_{i\in S}|\mbox{\boldmath  $W$}_{i\not\in S}^{(N)}).
\label{FanoInequality1}
\end{eqnarray}

For large enough $N$, (\ref{FanoInequality1}) implies that
\begin{eqnarray}
&& \sum_{i \in S} \tilde{r}_i \le \frac{1}{N}I(\mbox{\boldmath  $W$}^{(N)}_{i\in S} ; \hat{\mbox{\boldmath  $W$}}^{(N)}_{i\in S}|\mbox{\boldmath  $W$}_{i\not\in S}^{(N)})+  u_{\delta} \nonumber \\
&& \le \frac{1}{N}I_{\mbox{\scriptsize \boldmath  $r$}(\mbox{\scriptsize \boldmath  $W$}^{(N)})}(\mbox{\boldmath  $X$}^{(N)}_{i\in S} ; Y^{(N)}|\mbox{\boldmath  $X$}^{(N)}_{i\not\in S})+ u_{\delta} \nonumber \\
&& \le \frac{1}{N}H_{\mbox{\scriptsize \boldmath  $r$}(\mbox{\scriptsize \boldmath  $W$}^{(N)})}(Y^{(N)}|\mbox{\boldmath  $X$}^{(N)}_{i\not\in S})  - \frac{1}{N}H_{\mbox{\scriptsize \boldmath  $r$}(\mbox{\scriptsize \boldmath  $W$}^{(N)})}(Y^{(N)}|\mbox{\boldmath  $X$}^{(N)})+ u_{\delta} \nonumber \\
&& \le H_{\mbox{\scriptsize \boldmath  $r$}(\mbox{\scriptsize \boldmath  $W$}^{(N)})}(Y|\mbox{\boldmath  $X$}_{i\not\in S}) - H_{\mbox{\scriptsize \boldmath  $r$}(\mbox{\scriptsize \boldmath  $W$}^{(N)})}(Y|\mbox{\boldmath  $X$})+ u_{\delta} \nonumber \\
&& \le I_{\tilde{\mbox{\scriptsize \boldmath  $r$}}}(\mbox{\boldmath  $X$}_{i\in S}; Y|\mbox{\boldmath  $X$}_{i\not\in S})+ 2u_{\delta},
\label{FanoInequality2}
\end{eqnarray}
where the second inequality is due to the ``data processing inequality" \cite{ref Cover05}. Taking $\delta\to 0$ in (\ref{FanoInequality2}) yields (\ref{MutualInformationInequalityConverse}).
\end{proof}

\subsection{Proof of Theorem \ref{Theorem3}}
\label{ProofTheorem3}
\begin{proof}
Although the basic idea of the proof follows from Shannon's typical sequence arguments introduced in the channel capacity proof \cite{ref Shannon48}, it is significantly complicated by the following two factors. First, the receiver is only interested in recovering a particular user's message rather than messages of all users. Second, the empirical input distribution of different codewords may be different. Due to these complications, and for easy understanding, we choose to present the full proof in five detailed steps.

{\bf Step 1 (The typical sequence decoder): } $\quad$
We first define a typical sequence decoder as follows. Consider a sequence of random coding schemes $\{(\mbox{\boldmath  ${\cal L}$}^{(N)}, \mbox{\boldmath  $\gamma$}^{(N)})\}$ following asymptotic input distribution $\mbox{\boldmath  $P$}_{\mbox{\scriptsize \boldmath  $X$}|\mbox{\scriptsize \boldmath  $r$}}$. For any given codeword length $N$, let $\mbox{\boldmath  $X$}^{(N)}$ be the randomly chosen codeword associated with message $\mbox{\boldmath  $W$}^{(N)}$. Let  $\mbox{\boldmath  $Y$}^{(N)}$ be the channel output sequence. Define $\mbox{\boldmath  $r$}(\mbox{\boldmath  $X$}^{(N)})=\mbox{\boldmath  $r$}(\mbox{\boldmath  $W$}^{(N)})=\frac{1}{N}\log_2(\mbox{\boldmath  $W$}^{(N)})$ as the standard rate vector corresponding to $\mbox{\boldmath  $W$}^{(N)}$ and $\mbox{\boldmath  $X$}^{(N)}$. To simplify the proof, we assume that the input distribution of the random coding scheme $(\mbox{\boldmath  ${\cal L}$}^{(N)}, \mbox{\boldmath  $\gamma$}^{(N)})$, denoted by $\mbox{\boldmath  $P$}_{\mbox{\scriptsize \boldmath  $X$}|\mbox{\scriptsize \boldmath  $W$}^{(N)}}$, is indeed equal to the asymptotic input distribution $\mbox{\boldmath  $P$}_{\mbox{\scriptsize \boldmath  $X$}|\mbox{\scriptsize \boldmath  $r$}(\mbox{\scriptsize \boldmath  $W$}^{(N)})}$, for all $N$. Extending the proof to the general case is straightforward.

As in \cite{ref Berger77}\cite{ref Csiszar81}, we define the set $A^{(N)}_{\epsilon} (\mbox{\boldmath  $r$})$ of strongly typical sequences $\{(\mbox{\boldmath  $X$}^{(N)}, Y^{(N)})\}$ with respect to the distribution $P_{(\mbox{\scriptsize \boldmath  $X$}, Y)|\mbox{\scriptsize \boldmath  $r$}}= P_{Y|\mbox{\scriptsize \boldmath  $X$}} \prod_i P_{X_i|r_i}$ as the set of length-$N$ sequences whose empirical point mass functions are point-wise $\epsilon$-close to the true point mass function. Namely,
\begin{eqnarray}
&& A^{(N)}_{\epsilon} (\mbox{\boldmath  $r$})=\left\{ (\mbox{\boldmath  $X$}^{(N)}, Y^{(N)})\in \mbox{\boldmath  ${\cal X}$}^{N}\times {\cal Y}^{N} \right| \forall (\mbox{\boldmath  $X$}, Y)\in \mbox{\boldmath  ${\cal X}$}\times {\cal Y} \mbox{ with } P_{Y|\mbox{\scriptsize \boldmath  $X$}} \prod_i P_{X_i|r_i}>0 \nonumber \\
&& \left. \mbox{and we have } \left| \frac{1}{N} {\cal N}(\mbox{\boldmath  $X$}, Y|\mbox{\boldmath  $X$}^{(N)}, Y^{(N)}) - P_{Y|\mbox{\scriptsize \boldmath  $X$}} \prod_i P_{X_i|r_i} \right| < \frac{\epsilon}{ |{\cal Y}|\prod_i |{\cal X}_i|} \right\},
\label{TypicalSequenceSet}
\end{eqnarray}
where ${\cal N}(\mbox{\boldmath  $X$}, Y|\mbox{\boldmath  $X$}^{(N)}, Y^{(N)})$ is the number of occurrences of $(\mbox{\boldmath  $X$}, Y)$ in $(\mbox{\boldmath  $X$}^{(N)}, Y^{(N)})$.

Given a sequence pair $(\mbox{\boldmath  $X$}^{(N)}, Y^{(N)})$, we define $H_{(\mbox{\scriptsize \boldmath  $X$}^{(N)}, Y^{(N)})}( )$ as the entropy function computed using the empirical distribution of $(\mbox{\boldmath  $X$}^{(N)}, Y^{(N)})$. Given $\epsilon$, we can find a constant $\epsilon_1$, such that for all $(\mbox{\boldmath  $X$}^{(N)}, Y^{(N)})\in A^{(N)}_{\epsilon} (\mbox{\boldmath  $r$}(\mbox{\boldmath  $X$}^{(N)}))$ and for all user subset $S\subseteq \{1, \dots, K\}$, the following inequalities hold.
\begin{eqnarray}
&& |H_{(\mbox{\scriptsize \boldmath  $X$}^{(N)}, Y^{(N)})}(\mbox{\boldmath  $X$}_{i\in S}, Y) -H_{\mbox{\scriptsize \boldmath  $r$}(\mbox{\scriptsize \boldmath  $X$}^{(N)})}(\mbox{\boldmath  $X$}_{i\in S}, Y)| < \epsilon_1 \nonumber \\
&& |H_{(\mbox{\scriptsize \boldmath  $X$}^{(N)}, Y^{(N)})}(\mbox{\boldmath  $X$}_{i\in S}) -H_{\mbox{\scriptsize \boldmath  $r$}(\mbox{\scriptsize \boldmath  $X$}^{(N)})}(\mbox{\boldmath  $X$}_{i\in S})| < \epsilon_1.
\label{TypicalEntropyBound}
\end{eqnarray}
According to the strong typicality property (\ref{TypicalSequenceSet}), we can choose $\epsilon_1$ to satisfy $\epsilon_1\to 0$ as $N\to \infty$ and $\epsilon \to 0$,

Assume that, after observing the channel output sequence $Y^{(N)}$, the receiver constructs a set $X^{(N)}_{\epsilon}$ of codewords that are jointly typical with $Y^{(N)}$; that is,
\begin{equation}
X^{(N)}_{\epsilon}=\left\{\mbox{\boldmath  $X$}^{(N)}\left| \begin{array}{l} (\mbox{\boldmath  $X$}^{(N)}, Y^{(N)}) \in A^{(N)}_{\epsilon} (\mbox{\boldmath  $r$}),  \mbox{\boldmath  $r$}=\mbox{\boldmath  $r$}(\mbox{\boldmath  $X$}^{(N)}) \in \mbox{\boldmath  $R$}_{k} \end{array} \right. \right\}.
\end{equation}
The receiver outputs a message estimate $\hat{W}^{(N)}_k$ for user $k$ if, for all codewords $\mbox{\boldmath  $X$}^{(N)}\in A^{(N)}_{\epsilon}$, the $k$th codeword $X^{(N)}_k$ of $\mbox{\boldmath  $X$}^{(N)}\in A^{(N)}_{\epsilon}$ corresponds to the same message $\hat{W}^{(N)}_k$. If such $\hat{W}^{(N)}_k$ doesn't exist, the receiver outputs a collision for user $k$.

{\bf Step 2 (Some key definitions): } $\quad$
Let $\sigma>0$ be a small constant. We define $\mbox{\boldmath  $R$}_{k}^{(\sigma)}$ as a subset of $\mbox{\boldmath  $R$}_{k}$, by
\begin{equation}
\mbox{\boldmath  $R$}_{k}^{(\sigma)}=\left\{\mbox{\boldmath  $r$}\left| \begin{array}{l} \forall S \subseteq \{1, \dots, K\}, k\in S, \exists \tilde{S}\subseteq S, k\in \tilde{S}, \\ \mbox{such that}, \sum_{i\in \tilde{S}}r_i < I_{\mbox{\scriptsize \boldmath  $r$}} ( \mbox{\boldmath  $X$}_{i \in \tilde{S}} ; Y |\mbox{\boldmath  $X$}_{i \not\in S})-\sigma \end{array} \right.\right\}.
\label{MutualInformationInequality11}
\end{equation}

Let $M$ be a large integer whose value will be specified soon. We define $\left\{0, \frac{R_{\max}}{M}, \frac{2 R_{\max}}{M}, \dots, R_{\max}\right\}$ as the set of {\it grid rates} of a user. Define $r_g(W^{(N)})= \left\lfloor \frac{M r(W^{(N)})}{R_{\max}} \right\rfloor \frac{R_{\max}}{M} $ as a function that outputs the largest grid rate less than or equal to $r(W^{(N)})$. Let $\mbox{\boldmath  $r$}_g(\mbox{\boldmath  $W$}^{(N)})$ be the corresponding vector function of $r_g(W^{(N)})$.

Let $\{\mbox{\boldmath  $W$}^{(N)}\}$ be an arbitrary message sequence satisfying $\lim_{N\to \infty} \mbox{\boldmath  $r$}(\mbox{\boldmath  $W$}^{(N)})=\hat{\mbox{\boldmath  $r$}}$, for some $\hat{\mbox{\boldmath  $r$}}$. Assume that $\mbox{\boldmath  $W$}^{(N)}$ is the transmitted message with codeword length $N$. Let the actual codeword associated with $\mbox{\boldmath  $W$}^{(N)}$ be $\mbox{\boldmath  $X$}^{(N)}$, which is a random variable depending on the codebook index $\mbox{\boldmath  $\theta$}$. Let the channel output sequence be $Y^{(N)}$. It is easy to show that
\begin{equation}
\lim_{N\to \infty} Pr\left\{(\mbox{\boldmath  $X$}^{(N)}, Y^{(N)}) \in A^{(N)}_{\epsilon} \left(P_{(\mbox{\scriptsize \boldmath  $X$}, Y)|\mbox{\scriptsize \boldmath  $r$}(\mbox{\scriptsize \boldmath  $X$}^{(N)})}\right)\right\} =1.
\label{AsymptoticDetectability2}
\end{equation}
That is, with an asymptotic probability of one, the receiver will find $Y^{(N)}$ to be jointly typical with the transmitted codeword vector $\mbox{\boldmath  $X$}^{(N)}$.

{\bf Step 3 (Error type I, outputting the wrong message): } $\quad$
In this step, we will show the probability that the receiver can find another codeword vector $\tilde{\mbox{\boldmath  $W$}}^{(N)}$ in the achievable rate region, i.e., $\mbox{\boldmath  $r$}(\tilde{\mbox{\boldmath  $W$}}^{(N)})\in \mbox{\boldmath  $R$}_{k}$, with $\tilde{W}^{(N)}_k\ne W^{(N)}_k$, is asymptotically zero.

Without loss of generality, we focus our discussion on an arbitrary subset of messages in the achievable rate region. The subset is denoted by $\bar{B}^{(N)}_{S, \mathring{\mbox{\scriptsize \boldmath  $r$}}}$, illustrated in Figure \ref{Fig1}, with the notation being explained as follows. Superscript $N$ denotes the codeword length. Subscript $S$ is a subset of users $S\subseteq \{1, \dots, K\}$ with $k\in S$. We assume that, for all messages $\tilde{\mbox{\boldmath  $W$}}^{(N)}\in \bar{B}^{(N)}_{S, \mathring{\mbox{\scriptsize \boldmath  $r$}}}$ in the subset, we have $\tilde{\mbox{\boldmath  $W$}}^{(N)}_{i\not\in S}=\mbox{\boldmath  $W$}^{(N)}_{i\not\in S}$, and $\tilde{W}^{(N)}_i \neq W^{(N)}_i$, $\forall i\in S$. In other words, for all users not in $S$, their messages should equal to the corresponding transmitted messages, and for any user in $S$, its message should not equal to the transmitted message. Subscript $\mathring{\mbox{\boldmath  $r$}}$ is a vector characterizing the rates of messages in the subset. For all messages $\tilde{\mbox{\boldmath  $W$}}^{(N)}\in \bar{B}^{(N)}_{S, \mathring{\mbox{\scriptsize \boldmath  $r$}}}$, we have $\mbox{\boldmath  $r$}(\tilde{\mbox{\boldmath  $W$}}^{(N)}_{i\not\in S})=\mathring{\mbox{\boldmath  $r$}}_{i\not\in S}$, and $\mbox{\boldmath  $r$}_g(\tilde{\mbox{\boldmath  $W$}}^{(N)}_{i\in S})=\mathring{\mbox{\boldmath  $r$}}_{i\in S}$. In other words, entries of $\mathring{\mbox{\boldmath  $r$}}_{i\in S}$ take grid rate values, with $\mathring{\mbox{\boldmath  $r$}}_{i\in S} \le \mbox{\boldmath  $r$}(\tilde{\mbox{\boldmath  $W$}}^{(N)}_{i\in S}) < \mathring{\mbox{\boldmath  $r$}}_{i\in S}+ \frac{R_{\max}}{M} \mbox{\boldmath  $1$}_{i\in S}$. In summary, the message subset $\bar{B}^{(N)}_{S, \mathring{\mbox{\scriptsize \boldmath  $r$}}}$ is defined as follows.
\begin{equation}
\bar{B}^{(N)}_{S, \mathring{\mbox{\scriptsize \boldmath  $r$}}}=\left\{ \tilde{\mbox{\boldmath  $W$}}^{(N)} \left| \begin{array}{l} \tilde{\mbox{\boldmath  $W$}}^{(N)}_{i\not\in S}=\mbox{\boldmath  $W$}^{(N)}_{i\not\in S} \\ \tilde{W}^{(N)}_i \neq W^{(N)}_i, \forall i\in S \\ \mbox{\boldmath  $r$}(\tilde{\mbox{\boldmath  $W$}}^{(N)}_{i\not\in S})=\mathring{\mbox{\boldmath  $r$}}_{i\not\in S} \\ \mbox{\boldmath  $r$}_g(\tilde{\mbox{\boldmath  $W$}}^{(N)}_{i\in S})=\mathring{\mbox{\boldmath  $r$}}_{i\in S} \end{array} \right. \right\}.
\end{equation}

\begin{figure}[h]
  \centering
  \includegraphics[width=5 in]{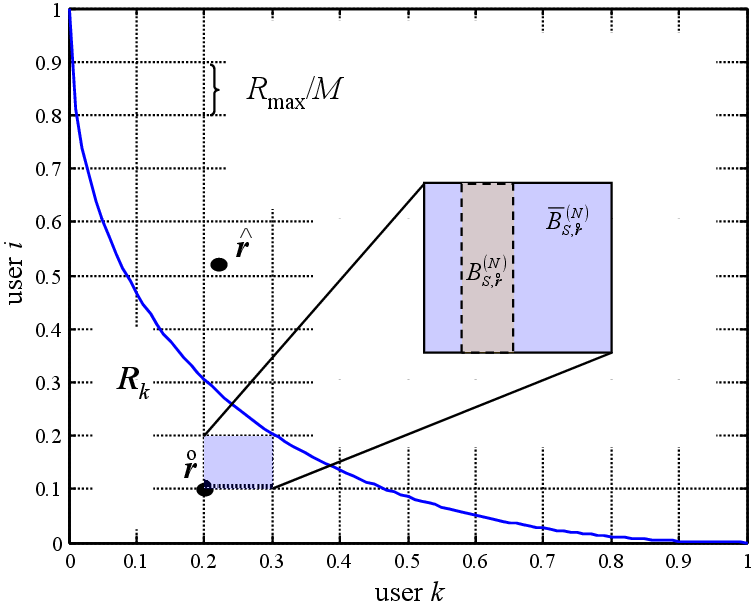}\caption{\label{Fig1} An illustration of achievable rate region and grid rates.}
\end{figure}

Let $M$ be large enough, such that the following inequality is satisfied for all messages $\tilde{\mbox{\boldmath  $W$}}^{(N)}\in \bar{B}^{(N)}_{S, \mathring{\mbox{\scriptsize \boldmath  $r$}}} $ and for all user subset $\breve{S} \subseteq \{1, \dots, K\}$:
\begin{eqnarray}
&& \left|H_{\mbox{\scriptsize \boldmath  $r$}(\tilde{\mbox{\scriptsize \boldmath  $W$}}^{(N)})}( \mbox{\boldmath  $X$}_{i\in \breve{S}} ; Y )              - H_{\mathring{\mbox{\scriptsize \boldmath  $r$}}}( \mbox{\boldmath  $X$}_{i\in \breve{S}} ; Y) \right| \le \epsilon_2, \nonumber \\
&& \left|H_{\mbox{\scriptsize \boldmath  $r$}(\tilde{\mbox{\scriptsize \boldmath  $W$}}^{(N)})}( \mbox{\boldmath  $X$}_{i\in \breve{S}})              - H_{\mathring{\mbox{\scriptsize \boldmath  $r$}}}( \mbox{\boldmath  $X$}_{i\in \breve{S}}) \right| \le \epsilon_2,
\label{GridRateBound}
\end{eqnarray}
where $\epsilon_2$ is a constant that satisfies $\epsilon_2\to 0$ as $M\to \infty$\footnote{Note that, if the asymptotic input distribution $\mbox{\boldmath  $P$}_{\mbox{\scriptsize \boldmath  $X$}|\mbox{\scriptsize \boldmath  $r$}}$ is not continuous in $\mbox{\boldmath  $r$}_{i\in S}$ at $\tilde{\mbox{\boldmath  $r$}}$ for all $\tilde{\mbox{\boldmath  $r$}}$ satisfying $\tilde{\mbox{\boldmath  $r$}}_{i\not\in S}=\mathring{\mbox{\boldmath  $r$}}_{i\not\in S}$, $\mathring{\mbox{\boldmath  $r$}}_{i\in S} \le \tilde{\mbox{\boldmath  $r$}}_{i\in S} < \mathring{\mbox{\boldmath  $r$}}_{i\in S}+ \frac{R_{\max}}{M} \mbox{\boldmath  $1$}_{i\in S}$, then (\ref{GridRateBound}) may not be satisfied only by setting $M$ at a large value. In this case, one needs to further partition the set $\bar{B}^{(N)}_{S, \mathring{\mbox{\scriptsize \boldmath  $r$}}}$ into a finite number of subsets, such that (\ref{GridRateBound}) is satisfied by messages within each subset. The partitioning is possible because the asymptotic input distribution of every user is discontinuous at most at a finite number of rate points. The rest of the proof can essentially be applied to each subset of messages with only minor revisions. Hence the corresponding detailed discussion is omitted.}. We also let $M$ be large enough such that
\begin{equation}
\frac{K R_{\max}}{M} \le \epsilon_3,
\label{MBound}
\end{equation}
where $\epsilon_3\to 0$ as $M\to \infty$.

Define $P_e^{(N)}\{\bar{B}^{(N)}_{S, \mathring{\mbox{\scriptsize \boldmath  $r$}}} \}$ as the probability that the receiver can find a message $\tilde{\mbox{\boldmath  $W$}}^{(N)} \in \bar{B}^{(N)}_{S, \mathring{\mbox{\scriptsize \boldmath  $r$}}}$ whose codeword $\tilde{\mbox{\boldmath  $X$}}^{(N)}$ satisfies $(\tilde{\mbox{\boldmath  $X$}}^{(N)}, Y^{(N)}) \in A^{(N)}_{\epsilon} \left(P_{(\mbox{\scriptsize \boldmath  $X$}, Y)|\mbox{\scriptsize \boldmath  $r$}(\tilde{\mbox{\scriptsize \boldmath  $X$}}^{(N)})}\right)$. We will show next that $\lim_{N\to \infty} P_e^{(N)}\{\bar{B}^{(N)}_{S, \mathring{\mbox{\scriptsize \boldmath  $r$}}} \} =0$, for all $\hat{\mbox{\boldmath  $r$}}$, $S$, and $\mathring{\mbox{\boldmath  $r$}}$ that satisfy the assumptions.

Since $\mathring{\mbox{\boldmath  $r$}} \in \mbox{\boldmath  $R$}_{k}^{(\sigma)}$, we can find a user subset $\tilde{S} \subseteq S$ with $k \in \tilde{S}$, such that
\begin{equation}
\sum_{i\in \tilde{S}}\mathring{r}_i < I_{\mathring{\mbox{\scriptsize \boldmath  $r$}}} ( \mbox{\boldmath  $X$}_{i \in \tilde{S}} ; Y |\mbox{\boldmath  $X$}_{i \not\in S})-\sigma.
\label{tilderandr}
\end{equation}
Let $\tilde{\mbox{\boldmath  $W$}}^{(N)} \in \bar{B}^{(N)}_{S, \mathring{\mbox{\scriptsize \boldmath  $r$}}}$ be a message whose codeword is denoted by $\tilde{\mbox{\boldmath  $X$}}^{(N)}$. Define $\tilde{\mbox{\boldmath  $r$}}=\mbox{\boldmath  $r$}(\tilde{\mbox{\boldmath  $W$}}^{(N)})$. We say $\tilde{\mbox{\boldmath  $X$}}^{(N)}_{i\in \tilde{S}}$ is jointly typical with $Y^{(N)}$ {\it with respect to} $\bar{B}^{(N)}_{S, \mathring{\mbox{\scriptsize \boldmath  $r$}}}$, denoted by $(\tilde{\mbox{\boldmath  $X$}}^{(N)}_{i\in \tilde{S}}, Y^{(N)}) \in A^{(N)}_{\epsilon} \left(P_{(\mbox{\scriptsize \boldmath  $X$}, Y)|\tilde{\mbox{\scriptsize \boldmath  $r$}}_{i\in \tilde{S}}, \bar{B}^{(N)}_{S, \mathring{\mbox{\scriptsize \boldmath  $r$}}}}\right)$, if there exists a codeword $\breve{\mbox{\boldmath  $X$}}^{(N)}$ with its corresponding message $\breve{\mbox{\boldmath  $W$}}^{(N)}$ that satisfies $\breve{\mbox{\boldmath  $W$}}^{(N)}\in \bar{B}^{(N)}_{S, \mathring{\mbox{\scriptsize \boldmath  $r$}}}$, $(\breve{\mbox{\boldmath  $X$}}^{(N)}, Y^{(N)}) \in A^{(N)}_{\epsilon} \left(P_{(\mbox{\scriptsize \boldmath  $X$}, Y)|\mbox{\scriptsize \boldmath  $r$}(\breve{\mbox{\scriptsize \boldmath  $X$}}^{(N)})}\right)$ and $\breve{\mbox{\boldmath  $X$}}^{(N)}_{i\in \tilde{S}}=\tilde{\mbox{\boldmath  $X$}}^{(N)}_{i\in \tilde{S}}$. According to the definition of $\bar{B}^{(N)}_{S, \mathring{\mbox{\scriptsize \boldmath  $r$}}}$, for all $i\in S$, we have $\tilde{W}^{(N)}_i\ne W^{(N)}_i$. Under this condition, $\tilde{\mbox{\boldmath  $X$}}^{(N)}_{i\in \tilde{S}}$ and $Y^{(N)}$ are generated independently. Consequently, due to the strongly typicality property (\ref{TypicalSequenceSet}), and inequalities (\ref{TypicalEntropyBound}), (\ref{GridRateBound}), the probability that $(\tilde{\mbox{\boldmath  $X$}}^{(N)}_{i\in \tilde{S}}, Y^{(N)}) \in A^{(N)}_{\epsilon} \left(P_{(\mbox{\scriptsize \boldmath  $X$}, Y)|\tilde{\mbox{\scriptsize \boldmath  $r$}}_{i\in \tilde{S}}, \bar{B}^{(N)}_{S, \mathring{\mbox{\scriptsize \boldmath  $r$}}} }\right)$ can be upper bounded as follows:
\begin{eqnarray}
&& \log_2\left[Pr\left\{(\tilde{\mbox{\boldmath  $X$}}^{(N)}_{i\in \tilde{S}}, Y^{(N)}) \in A^{(N)}_{\epsilon} \left(P_{(\mbox{\scriptsize \boldmath  $X$}, Y)|\tilde{\mbox{\scriptsize \boldmath  $r$}}_{i\in \tilde{S}}, \bar{B}^{(N)}_{S, \mathring{\mbox{\scriptsize \boldmath  $r$}}}}\right)\right\}\right] \nonumber \\
&& \le N\left[(H_{\mathring{\mbox{\scriptsize \boldmath  $r$}}}(\mbox{\boldmath  $X$}_{i\in \tilde{S}}, \mbox{\boldmath  $X$}_{i\not\in S}, Y)+\epsilon_1+\epsilon_2)    -(H_{\mathring{\mbox{\scriptsize \boldmath  $r$}}}(\mbox{\boldmath  $X$}_{i\in \tilde{S}})-\epsilon_1-\epsilon_2)          -(H_{\mathring{\mbox{\scriptsize \boldmath  $r$}}}(\mbox{\boldmath  $X$}_{i\not\in S}, Y)-\epsilon_1-\epsilon_2)\right] \nonumber \\
&& =-N\left[I_{\mathring{\mbox{\scriptsize \boldmath  $r$}}}(\mbox{\boldmath  $X$}_{i\in \tilde{S}}; \mbox{\boldmath  $X$}_{i\not\in S}, Y )-3\epsilon_1-3\epsilon_2\right] \nonumber \\
&& = -N\left[I_{\mathring{\mbox{\scriptsize \boldmath  $r$}}}(\mbox{\boldmath  $X$}_{i\in \tilde{S}}; Y | \mbox{\boldmath  $X$}_{i\not\in S} )-3\epsilon_1-3\epsilon_2 \right] .
\label{ErrorProbabilitybyMutualInformation}
\end{eqnarray}
Therefore,
\begin{eqnarray}
&& P_e^{(N)}\{\bar{B}^{(N)}_{S, \mathring{\mbox{\scriptsize \boldmath  $r$}}} \} \le   \sum_{\scriptsize  \tilde{\mbox{\scriptsize \boldmath  $W$}}^{(N)}_{i\in \tilde{S}}, \tilde{\mbox{\scriptsize \boldmath  $W$}}^{(N)}\in \bar{B}^{(N)}_{S, \mathring{\mbox{\tiny \boldmath  $r$}}}  } 2^{-N \left[I_{\mathring{\mbox{\tiny \boldmath  $r$}}}(\mbox{\scriptsize \boldmath  $X$}_{i\in \tilde{S}}; Y |\mbox{\scriptsize \boldmath  $X$}_{i\not\in S} ) -3\epsilon_1-3\epsilon_2 \right]}.
\label{JointlyTypicalProb}
\end{eqnarray}
Assume that $N$ is large enough to yield
\begin{equation}
\sum_{\scriptsize \tilde{\mbox{\scriptsize \boldmath  $W$}}^{(N)}_{i\in \tilde{S}}, \tilde{\mbox{\scriptsize \boldmath  $W$}}^{(N)}\in \bar{B}^{(N)}_{S, \mathring{\mbox{\tiny \boldmath  $r$}}}  } 1 < 2^{N \sum_{i\in \tilde{S}}(\mathring{r}_i+\frac{R_{\max}}{M})} \le  2^{N\left( \sum_{i\in \tilde{S}}\mathring{r}_i+\frac{KR_{\max}}{M}\right)}\le 2^{N\left( \sum_{i\in \tilde{S}}\mathring{r}_i+\epsilon_3\right)},
\label{LargeCardinality}
\end{equation}
where the first inequality is due to the fact that $\tilde{\mbox{\boldmath  $W$}}^{(N)}\in \bar{B}^{(N)}_{S, \mathring{\mbox{\scriptsize \boldmath  $r$}}}$ implies that $r(\tilde{W}^{(N)}_i)\le \mathring{r}_i+\frac{R_{\max}}{M}$ for all $i\in \tilde{S}$, and the last inequality is due to (\ref{MBound}).

Consequently, (\ref{JointlyTypicalProb}) and (\ref{LargeCardinality}) lead to
\begin{eqnarray}
&& \log_2\left(P_e^{(N)}\{\bar{B}^{(N)}_{S, \mathring{\mbox{\scriptsize \boldmath  $r$}}} \}\right)  < N\left[ \sum_{i\in \tilde{S}}\mathring{r}_i -I_{\mathring{\mbox{\scriptsize \boldmath  $r$}}}(\mbox{\boldmath  $X$}_{i\in \tilde{S}}; Y |\mbox{\boldmath  $X$}_{i\not\in S} ) + 3\epsilon_1+3\epsilon_2+\epsilon_3 \right].
\label{JointlyTypicalProb2}
\end{eqnarray}
Because $\sum_{i\in \tilde{S}}\mathring{r}_i - I_{\mathring{\mbox{\scriptsize \boldmath  $r$}}} ( \mbox{\boldmath  $X$}_{i \in \tilde{S}} ; Y |\mbox{\boldmath  $X$}_{i \not\in S})+\sigma <0$, choose $3\epsilon_1+3\epsilon_2+\epsilon_3 < \sigma$, we get
\begin{equation}
\lim_{N\to \infty} P_e^{(N)}\{\bar{B}^{(N)}_{S, \mathring{\mbox{\scriptsize \boldmath  $r$}}} \} =0.
\label{PeGoestoZero}
\end{equation}

Note that, given $\sigma$, (\ref{PeGoestoZero}) holds for all $\hat{\mbox{\boldmath  $r$}}$, $S$, and $\mathring{\mbox{\boldmath  $r$}}$ satisfying our assumptions. Therefore, by taking $M\to \infty$, we can see the probability that the receiver finds a message $\tilde{\mbox{\boldmath  $W$}}^{(N)}$ in $\mbox{\boldmath  $R$}_{k}^{(\sigma)}$ with codeword $\tilde{\mbox{\boldmath  $X$}}^{(N)}$ being jointly typical with $Y^{(N)}$ and $\tilde{W}_k^{(N)}\ne W_k^{(N)}$ is asymptotically zero. Next, by taking $\sigma \to 0$, we conclude that, for all message sequences $\{\mbox{\boldmath  $W$}^{(N)}\}$, the probability for the receiver to output $\tilde{W}_k^{(N)}\ne W_k^{(N)}$ is asymptotically zero. As a special case, for all message sequences $\{\mbox{\boldmath  $W$}^{(N)}\}$ with $\mbox{\boldmath  $r$}(\mbox{\boldmath  $W$}^{(N)})\in \mbox{\boldmath  $R$}_k$, $\forall N$, and $\lim_{N\to\infty}\mbox{\boldmath  $r$}(\mbox{\boldmath  $W$}^{(N)})\in \mbox{\boldmath  $R$}_k$, we have $\lim_{N\to \infty} P_{ek}(\mbox{\boldmath  $W$}^{(N)})=0$, where $P_{ek}(\mbox{\boldmath  $W$}^{(N)})$ is defined in (\ref{PekPck}).

{\bf Step 4 (Error type II, failing to report a collision): } $\quad$ In this step, we will show that if the standard rate vector $\mbox{\boldmath  $r$}(\mbox{\boldmath  $X$}^{(N)})$ is outside the achievable rate region, i.e., $\mbox{\boldmath  $r$}(\mbox{\boldmath  $X$}^{(N)})\not\in \mbox{\boldmath  $R$}_{kc}$, then the probability that the receiver does not report a collision is asymptotically zero. Based on the result that we have already demonstrated in Step 3, we only need to show that, asymptotically, the receiver is not able to find another codeword vector $\tilde{\mbox{\boldmath  $X$}}^{(N)}$ inside the rate region, $\mbox{\boldmath  $r$}(\tilde{\mbox{\boldmath  $X$}}^{(N)})\in \mbox{\boldmath  $R$}_{k}$, with $\tilde{X}^{(N)}_k= X^{(N)}_k$ and $\tilde{\mbox{\boldmath  $X$}}^{(N)}$ being jointly typical with $Y^{(N)}$.

We again focus our discussion on a message subset denoted by $B^{(N)}_{S, \mathring{\mbox{\scriptsize \boldmath  $r$}}}$, as illustrated in Figure \ref{Fig1}, with the notation being explained below. As before, superscript $N$ denotes the codeword length. Subscript $S$ is a user subset $S\subseteq \{1, \dots, K\}$ with $k\in S$. Also as in Step 3, we assume that, for all messages $\tilde{\mbox{\boldmath  $W$}}^{(N)} \in B^{(N)}_{S, \mathring{\mbox{\scriptsize \boldmath  $r$}}}$ in the set, we have $\tilde{\mbox{\boldmath  $W$}}^{(N)}_{i\not\in S}=\mbox{\boldmath  $W$}^{(N)}_{i\not\in S}$, and for all $i\in S \setminus \{k\}$, we have $\tilde{W}^{(N)}_i\neq W^{(N)}_i$. However, by contrast to Step 3, we assume that $\tilde{W}^{(N)}_k= W^{(N)}_k$ \footnote{Note that in Step 3, we already showed the probability for the receiver to output an erroneous message estimate is asymptotically zero. Therefore, in this Step, we only consider the possible situation that the receiver outputs the correct message.}. Subscript $\mathring{\mbox{\boldmath  $r$}}$ is a vector characterizing the rates of messages in the subset. For all messages $\tilde{\mbox{\boldmath  $W$}}^{(N)}\in B^{(N)}_{S, \mathring{\mbox{\scriptsize \boldmath  $r$}}}$, we have $\mbox{\boldmath  $r$}(\tilde{\mbox{\boldmath  $W$}}^{(N)}_{i\not\in S\setminus \{k \}})=\mathring{\mbox{\boldmath  $r$}}_{i\not\in S\setminus \{k \}}$, and $\mbox{\boldmath  $r$}_g(\tilde{\mbox{\boldmath  $W$}}^{(N)}_{i\in S \setminus \{k \}})=\mathring{\mbox{\boldmath  $r$}}_{i\in S \setminus \{k \}}$. In other words, entries of $\mathring{\mbox{\boldmath  $r$}}_{i\in S \setminus \{k \}}$ take grid rate values, with $\mathring{\mbox{\boldmath  $r$}}_{i\in S \setminus \{k \}} \le \mbox{\boldmath  $r$}(\tilde{\mbox{\boldmath  $W$}}^{(N)}_{i\in S \setminus \{k \}}) < \mathring{\mbox{\boldmath  $r$}}_{i\in S \setminus \{k \}}+ \frac{R_{\max}}{M} \mbox{\boldmath  $1$}_{i\in S \setminus \{k \}}$. In summary, the message set $B^{(N)}_{S, \mathring{\mbox{\scriptsize \boldmath  $r$}}}$ is defined by

\begin{equation}
B^{(N)}_{S, \mathring{\mbox{\scriptsize \boldmath  $r$}}}=\left\{ \tilde{\mbox{\boldmath  $W$}}^{(N)} \left| \begin{array}{l} \tilde{\mbox{\boldmath  $W$}}^{(N)}_{i\not\in S \setminus \{k\}}=\mbox{\boldmath  $W$}^{(N)}_{i\not\in S \setminus \{k\}} \\ \tilde{W}^{(N)}_i\neq W^{(N)}_i, \forall i \in S \setminus \{k\} \\ \mbox{\boldmath  $r$}(\tilde{\mbox{\boldmath  $W$}}^{(N)}_{i\not\in S\setminus \{k \}})=\mathring{\mbox{\boldmath  $r$}}_{i\not\in S\setminus \{k \}} \\ \mbox{\boldmath  $r$}_g(\tilde{\mbox{\boldmath  $W$}}^{(N)}_{i\in S \setminus \{k \}})=\mathring{\mbox{\boldmath  $r$}}_{i\in S \setminus \{k \}} \end{array} \right. \right\}.
\end{equation}

Define $\bar{P}_c^{(N)}\{B^{(N)}_{S, \mathring{\mbox{\scriptsize \boldmath  $r$}}} \}$ as the probability that the receiver can find a message $\tilde{\mbox{\boldmath  $W$}}^{(N)} \in B^{(N)}_{S, \mathring{\mbox{\scriptsize \boldmath  $r$}}}$ whose codeword $\tilde{\mbox{\boldmath  $X$}}^{(N)}$ satisfies $(\tilde{\mbox{\boldmath  $X$}}^{(N)}, Y^{(N)}) \in A^{(N)}_{\epsilon} \left(P_{(\mbox{\scriptsize \boldmath  $X$}, Y)|\mbox{\scriptsize \boldmath  $r$}(\tilde{\mbox{\scriptsize \boldmath  $X$}}^{(N)})}\right)$. We will show next that, for all $\hat{\mbox{\boldmath  $r$}} \not\in \mbox{\boldmath  $R$}_{kc}$, $S$, and $\mathring{\mbox{\boldmath  $r$}}$, $\lim_{N\to \infty} \bar{P}_c^{(N)}\{B^{(N)}_{S, \mathring{\mbox{\scriptsize \boldmath  $r$}}} \} =0$.

We let $M$ be large enough, such that inequality (\ref{GridRateBound}) is satisfied for all messages $\tilde{\mbox{\boldmath  $W$}}^{(N)}\in B^{(N)}_{S, \mathring{\mbox{\scriptsize \boldmath  $r$}}} $ and for all user subsets $\breve{S} \subseteq \{1, \dots, K\}$\footnote{As in Step 3, if the asymptotic input distribution $\mbox{\boldmath  $P$}_{\mbox{\scriptsize \boldmath  $X$}|\mbox{\scriptsize \boldmath  $r$}}$ is not continuous in $\mbox{\boldmath  $r$}_{i\in S\setminus \{k\}}$ at $\tilde{\mbox{\boldmath  $r$}}$ for all $\tilde{\mbox{\boldmath  $r$}}$ satisfying $\tilde{\mbox{\boldmath  $r$}}_{i\not\in S\setminus \{k\}}=\mathring{\mbox{\boldmath  $r$}}_{i\not\in S\setminus \{k\}}$, $\mathring{\mbox{\boldmath  $r$}}_{i\in S\setminus \{k\}} \le \tilde{\mbox{\boldmath  $r$}}_{i\in S\setminus \{k\}} < \mathring{\mbox{\boldmath  $r$}}_{i\in S\setminus \{k\}}+ \frac{R_{\max}}{M} \mbox{\boldmath  $1$}_{i\in S\setminus \{k\}}$, then (\ref{GridRateBound}) may not be satisfied only by setting $M$ at a large value. In this case, one needs to further partition the set $B^{(N)}_{S, \mathring{\mbox{\scriptsize \boldmath  $r$}}}$ into a finite number of subsets, such that (\ref{GridRateBound}) is satisfied by messages within each subset. The rest of the proof can be applied to each subset of messages with only minor revisions. Hence the corresponding detailed discussion is omitted.}. We also let $M$ be large enough, so that (\ref{MBound}) holds.

Next, we present a key proposition to support the rest of the proof.

\begin{proposition}{\label{Proposition2}}
There exists a user subset $S_1$ with $S_1 \cap S =\phi$, such that for all user subsets $S_2\subseteq \overline{S \cup S_1}$, with $\overline{S \cup S_1}$ being the complement of $S \cup S_1$, we have
\begin{equation}
\sum_{i\in S_2} \hat{r}_i + \hat{r}_k > I_{\hat{\mbox{\scriptsize \boldmath  $r$}}}(\mbox{\boldmath  $X$}_{i\in S_2\cup \{k\}} ; Y |\mbox{\boldmath  $X$}_{i\in S_1} ).
\label{RInequality1}
\end{equation}
Consequently, let $\mbox{\boldmath  $r$}=\mbox{\boldmath  $r$}(\mbox{\boldmath  $W$}^{(N)})$ be the standard rate of the actual transmitted message. For any positive constant $\epsilon_4$, we can find a large enough $N$ to satisfy
\begin{equation}
\sum_{i\in S_2} r_i + r_k > I_{\mbox{\scriptsize \boldmath  $r$}}(\mbox{\boldmath  $X$}_{i\in S_2\cup \{k\}} ; Y |\mbox{\boldmath  $X$}_{i\in S_1} )-\epsilon_4.
\label{RInequality1.5}
\end{equation}
$\QED$
\end{proposition}

The proof of Proposition \ref{Proposition2} is given in Appendix \ref{ProofProposition2}.

Let $S_1$ be the user subset found in Proposition \ref{Proposition2}. Let $\tilde{\mbox{\boldmath  $W$}}^{(N)} \in B^{(N)}_{S, \mathring{\mbox{\scriptsize \boldmath  $r$}}}$ be a message whose codeword is denoted by $\tilde{\mbox{\boldmath  $X$}}^{(N)}$. Define $\tilde{\mbox{\boldmath  $r$}}=\mbox{\boldmath  $r$}(\tilde{\mbox{\boldmath  $W$}}^{(N)})$. According to (\ref{AsymptoticDetectability2}), for a large enough $N$, the received signal $\mbox{\boldmath  $Y$}^{(N)}$ is jointly typical with the actual codeword $\mbox{\boldmath  $X$}^{(N)}$. Consequently, due to (\ref{TypicalEntropyBound}), for all user subsets $S_2\subseteq \overline{S \cup S_1}$, we have with high probability
\begin{equation}
\sum_{i\in S_2} r_i + r_k > I_{(\mbox{\scriptsize \boldmath  $X$}^{(N)}, Y^{(N)})}(\mbox{\boldmath  $X$}_{i\in S_2\cup \{k\}} ; Y |\mbox{\boldmath  $X$}_{i\in S_1} )-\epsilon_4-4\epsilon_1,
\label{ConditionalMutualInfo1}
\end{equation}
where $I_{(\mbox{\scriptsize \boldmath  $X$}^{(N)}, Y^{(N)})}(\mbox{\boldmath  $X$}_{i\in S_2\cup \{k\}} ; Y |\mbox{\boldmath  $X$}_{i\in S_1} )$ is the conditional mutual information computed using the empirical distribution of $(\mbox{\boldmath  $X$}^{(N)}, Y^{(N)})$.

Assume the receiver finds another message $\tilde{\mbox{\boldmath  $W$}}^{(N)} \in B^{(N)}_{S, \mathring{\mbox{\scriptsize \boldmath  $r$}}}$, whose codeword $\tilde{\mbox{\boldmath  $X$}}^{(N)}$ is jointly typical with $Y^{(N)}$. Since $\tilde{\mbox{\boldmath  $X$}}^{(N)}_{i\not\in S\setminus \{k\}}= \mbox{\boldmath  $X$}^{(N)}_{i\not\in S\setminus \{k\}}$, (\ref{ConditionalMutualInfo1}), (\ref{TypicalEntropyBound}) and (\ref{GridRateBound}) yield
\begin{eqnarray}
& \sum_{i\in S_2} \tilde{r}_i + \tilde{r}_k &= \sum_{i\in S_2} r_i + r_k > I_{(\mbox{\scriptsize \boldmath  $X$}^{(N)}, Y^{(N)})}(\mbox{\boldmath  $X$}_{i\in S_2\cup \{k\}} ; Y |\mbox{\boldmath  $X$}_{i\in S_1} )-\epsilon_4-4\epsilon_1 \nonumber \\
&& = I_{(\tilde{\mbox{\scriptsize \boldmath  $X$}}^{(N)}, Y^{(N)})}(\mbox{\boldmath  $X$}_{i\in S_2\cup \{k\}} ; Y |\mbox{\boldmath  $X$}_{i\in S_1} )-\epsilon_4-4\epsilon_1 \nonumber \\
&& > I_{\tilde{\mbox{\scriptsize \boldmath  $r$}}}(\mbox{\boldmath  $X$}_{i\in S_2\cup \{k\}} ; Y |\mbox{\boldmath  $X$}_{i\in S_1} )-\epsilon_4-8\epsilon_1 \nonumber \\
&& > I_{\mathring{\mbox{\scriptsize \boldmath  $r$}}}(\mbox{\boldmath  $X$}_{i\in S_2\cup \{k\}} ; Y |\mbox{\boldmath  $X$}_{i\in S_1} )-\epsilon_4-8\epsilon_1-4\epsilon_2.
\label{RInequality3}
\end{eqnarray}

By assumption, we have $\tilde{\mbox{\boldmath  $r$}}\in \mbox{\boldmath  $R$}_{k}^{(\sigma)} $. From (\ref{MutualInformationInequality11})\footnote{Here we regard $\bar{S}_1$, which is the compliment of $S_1$, as the user subset $S$ in (\ref{MutualInformationInequality11}).} and (\ref{GridRateBound}), we know that there exists a user subset $\tilde{S} \subseteq S$, $k\in \tilde{S}$, and a user subset $S_2\subseteq \overline{S \cup S_1}$, such that
\begin{equation}
\sum_{i\in S_2\cup \tilde{S}} \tilde{r}_i < I_{\tilde{\mbox{\scriptsize \boldmath  $r$}}}(\mbox{\boldmath  $X$}_{i\in S_2 \cup \tilde{S}} ; Y |\mbox{\boldmath  $X$}_{i\in S_1} )-\sigma < I_{\mathring{\mbox{\scriptsize \boldmath  $r$}}}(\mbox{\boldmath  $X$}_{i\in S_2 \cup \tilde{S}} ; Y |\mbox{\boldmath  $X$}_{i\in S_1} )+4\epsilon_2-\sigma.
\label{RInequality31}
\end{equation}
Combining (\ref{RInequality3}) and (\ref{RInequality31}), we obtain
\begin{eqnarray}
& \sum_{i\in \tilde{S}\setminus\{k\}} \tilde{r}_i & < I_{\mathring{\mbox{\scriptsize \boldmath  $r$}}}(\mbox{\boldmath  $X$}_{i\in \tilde{S}\setminus\{k\}} ; Y |\mbox{\boldmath  $X$}_{i\in S_1 \cup S_2\cup \{k\}} )-\sigma+8\epsilon_1+8\epsilon_2+\epsilon_4 \nonumber \\
&& \le I_{\mathring{\mbox{\scriptsize \boldmath  $r$}}}(\mbox{\boldmath  $X$}_{i\in \tilde{S}\setminus\{k\}} ; Y |\mbox{\boldmath  $X$}_{i\not\in S\setminus \{k\}} )-\sigma+8\epsilon_1+8\epsilon_2+\epsilon_4,
\label{RInequality32}
\end{eqnarray}
where the last inequality is due to the fact that codewords of different users are independent.

According to the definition of $B^{(N)}_{S, \mathring{\mbox{\scriptsize \boldmath  $r$}}}$, for all $i\in S\setminus \{k\}$, we have $\tilde{W}^{(N)}_i\ne W^{(N)}_i$. Under this condition, $\tilde{\mbox{\boldmath  $X$}}^{(N)}_{i\in \tilde{S}\setminus \{k\}}$ and $Y^{(N)}$ are independent. Consequently, the probability that the receiver finds $\tilde{\mbox{\boldmath  $X$}}^{(N)}_{i\in \tilde{S}\setminus \{k\}}$ being jointly typical with $Y^{(N)}$ {\it with respect to} $B^{(N)}_{S, \mathring{\mbox{\scriptsize \boldmath  $r$}}}$, denoted by $(\tilde{\mbox{\boldmath  $X$}}^{(N)}_{i\in \tilde{S}\setminus \{k\}}, Y^{(N)}) \in A^{(N)}_{\epsilon} \left(P_{(\mbox{\scriptsize \boldmath  $X$}, Y)|\tilde{\mbox{\scriptsize \boldmath  $r$}}_{i\in \tilde{S}\setminus \{k\}}, B^{(N)}_{S, \mathring{\mbox{\scriptsize \boldmath  $r$}}}}\right)$,  is upper bounded as follows:
\begin{eqnarray}
&& \log_2\left[Pr\left\{(\tilde{\mbox{\boldmath  $X$}}^{(N)}_{i\in \tilde{S}\setminus \{k\}}, Y^{(N)}) \in A^{(N)}_{\epsilon} \left(P_{(\mbox{\scriptsize \boldmath  $X$}, Y)|\tilde{\mbox{\scriptsize \boldmath  $r$}}_{i\in \tilde{S}\setminus \{k\}}, B^{(N)}_{S, \mathring{\mbox{\scriptsize \boldmath  $r$}}}}\right)\right\}\right] \nonumber \\
&& \le N\left[(H_{\mathring{\mbox{\scriptsize \boldmath  $r$}}}(\mbox{\boldmath  $X$}_{i\in \tilde{S}\setminus \{k\}}, \mbox{\boldmath  $X$}_{i\not\in S\setminus \{k\}}, Y )+\epsilon_1+\epsilon_2)     -(H_{\mathring{\mbox{\scriptsize \boldmath  $r$}}}(\mbox{\boldmath  $X$}_{i\in \tilde{S}\setminus \{k\}} )-\epsilon_1 -\epsilon_2) \right. \nonumber \\
&& \qquad  \left.  -(H_{\mathring{\mbox{\scriptsize \boldmath  $r$}}}(\mbox{\boldmath  $X$}_{i\not\in S\setminus \{k\}}, Y)-\epsilon_1-\epsilon_2)\right] \nonumber \\
&& = -N\left[I_{\mathring{\mbox{\scriptsize \boldmath  $r$}}}(\mbox{\boldmath  $X$}_{i\in \tilde{S}\setminus \{k\}}; \mbox{\boldmath  $X$}_{i\not\in S\setminus \{k\}}, Y )-3\epsilon_1 - 3\epsilon_2 \right] \nonumber \\
&& = -N\left[I_{\mathring{\mbox{\scriptsize \boldmath  $r$}}}(\mbox{\boldmath  $X$}_{i\in \tilde{S}\setminus \{k\}}; Y | \mbox{\boldmath  $X$}_{i\not\in S\setminus \{k\}})-3\epsilon_1 -3\epsilon_2 \right] .
\label{ErrorProbabilitybyMutualInformation2}
\end{eqnarray}
Therefore,
\begin{eqnarray}
&& \bar{P}_c^{(N)}\{B^{(N)}_{S, \mathring{\mbox{\scriptsize \boldmath  $r$}}} \}   \le  \sum_{\scriptsize \tilde{\mbox{\scriptsize \boldmath  $W$}}^{(N)}_{i\in \tilde{S}\setminus \{k\}}, \tilde{\mbox{\scriptsize \boldmath  $W$}}^{(N)}\in B^{(N)}_{S, \mathring{\mbox{\tiny \boldmath  $r$}}} } 2^{-N \left[I_{\mathring{\mbox{\tiny \boldmath  $r$}}}(\mbox{\scriptsize \boldmath  $X$}_{i\in \tilde{S}\setminus\{k\}}; Y |\mbox{\scriptsize \boldmath  $X$}_{i\not\in S\setminus\{k\}} ) -3\epsilon_1 -3\epsilon_2 \right]}.
\label{JointlyTypicalProb2}
\end{eqnarray}
Assume that $N$ is large enough to yield
\begin{eqnarray}
&& \sum_{\tilde{\mbox{\scriptsize \boldmath  $W$}}^{(N)}_{i\in \tilde{S}\setminus \{k\}}, \tilde{\mbox{\scriptsize \boldmath  $W$}}^{(N)}\in B^{(N)}_{S, \mathring{\mbox{\tiny \boldmath  $r$}}} }  1 < 2^{N \sum_{i\in \tilde{S}\setminus\{k\}}(\mathring{r}_i+\frac{R_{\max}}{M})}  \le  2^{N\left( \sum_{i\in \tilde{S}\setminus\{k\}}\mathring{r}_i+\frac{KR_{\max}}{M}\right)}\le 2^{N\left( \sum_{i\in \tilde{S}\setminus\{k\}}\mathring{r}_i+\epsilon_3\right)} .
\label{LargeCardinality2}
\end{eqnarray}
Consequently, (\ref{JointlyTypicalProb2}) and (\ref{LargeCardinality2}) lead to
\begin{eqnarray}
&& \log_2\left(\bar{P}_c^{(N)}\{B^{(N)}_{S, \mathring{\mbox{\scriptsize \boldmath  $r$}}} \}\right)  < N\left[ \sum_{i\in \tilde{S}\setminus\{k\}}\mathring{r}_i -I_{\mathring{\mbox{\scriptsize \boldmath  $r$}}}(\mbox{\boldmath  $X$}_{i\in \tilde{S}\setminus\{k\}}; Y |\mbox{\boldmath  $X$}_{i\not\in S\setminus\{k\}} ) + 3\epsilon_1+3\epsilon_2 +\epsilon_3\right].
\label{JointlyTypicalProb3}
\end{eqnarray}
If we choose $11\epsilon_1+11\epsilon_2+\epsilon_3+\epsilon_4 < \sigma$, we obtain from (\ref{RInequality32}) and (\ref{JointlyTypicalProb3}) that
\begin{equation}
\lim_{N\to \infty} \bar{P}_c^{(N)}\{B^{(N)}_{S, \mathring{\mbox{\scriptsize \boldmath  $r$}}} \} =0.
\label{PeGoestoZero2}
\end{equation}

Note that, given $\sigma$, (\ref{PeGoestoZero2}) holds for all $\hat{\mbox{\boldmath  $r$}}$, $S$, and $\mathring{\mbox{\boldmath  $r$}}$ satisfying the assumptions. Therefore, by taking $M\to \infty$, we can see that, if $\hat{\mbox{\boldmath  $r$}}\not\in \mbox{\boldmath  $R$}_{kc}$, then the probability that the receiver finds a message $\tilde{\mbox{\boldmath  $W$}}^{(N)}$ in $\mbox{\boldmath  $R$}_{k}^{(\sigma)}$ with codeword $\tilde{\mbox{\boldmath  $X$}}^{(N)}$ being jointly typical with $Y^{(N)}$ and $\tilde{W}_k^{(N)}= W_k^{(N)}$ is asymptotically zero. Next, by taking $\sigma \to 0$, we conclude that, for all message sequences $\{\mbox{\boldmath  $W$}^{(N)}\}$ with $\mbox{\boldmath  $r$}(\mbox{\boldmath  $W$}^{(N)})\not\in \mbox{\boldmath  $R$}_{kc}$, $\forall N$, and $\lim_{N\to\infty}\mbox{\boldmath  $r$}(\mbox{\boldmath  $W$}^{(N)})\not\in \mbox{\boldmath  $R$}_{kc}$, we have $\lim_{N\to \infty} P_{kc}(\mbox{\boldmath  $W$}^{(N)})=1$.

We have now proved that $\mbox{\boldmath  $R$}_{k}$ is asymptotically achievable for user $k$.
\end{proof}

\subsection{Proof of Proposition \ref{Proposition2}}
\label{ProofProposition2}
\begin{proof}
If the claim of the proposition is not true, then for all user subsets $S_1$ with $S_1 \cap S =\phi$, there exists a user subset $S_2\subseteq \overline{S \cup S_1}$ that satisfies
\begin{equation}
\sum_{i\in S_2} \hat{r}_i + \hat{r}_k \le I_{\hat{\mbox{\scriptsize \boldmath  $r$}}}(\mbox{\boldmath  $X$}_{i\in S_2\cup \{k\}} ; Y |\mbox{\boldmath  $X$}_{i\in S_1} ).
\label{RInequality6}
\end{equation}
Consequently, given the mutually exclusive user subsets $S$, $S_1$ and $S_2$ with $k \in S$, for all user subsets $S_3$ with $k\not\in S_3$ and $S_3\cap S_1=S_3\cap S_2=\phi$, we have
\begin{eqnarray}
&& \sum_{i\in S_2} \hat{r}_i + \hat{r}_k \le I_{\hat{\mbox{\scriptsize \boldmath  $r$}}}(\mbox{\boldmath  $X$}_{i\in S_2\cup \{k\}} ; Y |\mbox{\boldmath  $X$}_{i\in S_1} ) \nonumber \\
&& \quad = H_{\hat{\mbox{\scriptsize \boldmath  $r$}}}(\mbox{\boldmath  $X$}_{i\in S_2\cup \{k\}} ) -H_{\hat{\mbox{\scriptsize \boldmath  $r$}}}(\mbox{\boldmath  $X$}_{i\in S_2\cup \{k\}} | \mbox{\boldmath  $X$}_{i\in S_1}, Y ) \nonumber \\
&& \quad \le H_{\hat{\mbox{\scriptsize \boldmath  $r$}}}(\mbox{\boldmath  $X$}_{i\in S_2\cup \{k\}} ) -H_{\hat{\mbox{\scriptsize \boldmath  $r$}}}(\mbox{\boldmath  $X$}_{i\in S_2\cup \{k\}} | \mbox{\boldmath  $X$}_{i\in S_1 \cup S_3}, Y ) \nonumber \\
&& \quad =I_{\hat{\mbox{\scriptsize \boldmath  $r$}}}(\mbox{\boldmath  $X$}_{i\in S_2\cup \{k\}} ; Y |\mbox{\boldmath  $X$}_{i\in S_1 \cup S_3} ) .
\label{RInequality7}
\end{eqnarray}

From the derivation, we can see that $S_1 \cup S_3$ can be chosen to be any user subset satisfying $S_1 \cup S_3 \subseteq \{1, \dots, K\}\setminus \{k\}$. Consequently, (\ref{RInequality7}) implies $\hat{\mbox{\boldmath  $r$}} \in \mbox{\boldmath  $R$}_{kc}$, where $\mbox{\boldmath  $R$}_{kc}$ is the closure of $\mbox{\boldmath  $R$}_k$ defined in (\ref{MutualInformationInequality1}). This contradicts the assumption that $\hat{\mbox{\boldmath  $r$}} \not\in \mbox{\boldmath  $R$}_{kc}$. Therefore, the conclusion of the proposition must hold.
\end{proof}

\subsection{Proof of Theorem \ref{Theorem4}}
\label{ProofTheorem4}
\begin{proof}
Assume that we are given a set of rate regions $\{\mbox{\boldmath  $R$}_k, k \in S_0\}$, where $\mbox{\boldmath  $R$}_k$ is asymptotically achievable for user $k$. We will show next that if the regions $\mbox{\boldmath  $R$}_k$ are defined by (\ref{MutualInformationInequality1}) for $k\in S_0$, then $\mbox{\boldmath  $R$}_{S_0}$ defined by (\ref{MutualInformationInequality2}) is given by $\mbox{\boldmath  $R$}_{S_0}=\bigcap_{k\in S_0} \mbox{\boldmath  $R$}_k$. Since it is easy to see that $\mbox{\boldmath  $R$}_{S_0}\subseteq \bigcap_{k\in S_0} \mbox{\boldmath  $R$}_k$, we only need to show $\mbox{\boldmath  $R$}_{S_0}\supseteq \bigcap_{k\in S_0} \mbox{\boldmath  $R$}_k$, i.e., $\mbox{\boldmath  $r$}\in \mbox{\boldmath  $R$}_k$ for all $k\in S_0$ implies $\mbox{\boldmath  $r$} \in \mbox{\boldmath  $R$}_{S_0}$.

Assume $\mbox{\boldmath  $r$}\in \mbox{\boldmath  $R$}_k$ for all $k\in S_0$. Given a user subset $S \cap S_0 \ne \phi$, we can find a user subset $\tilde{S}_1 \subseteq S$ and $\tilde{S}_1 \cap S_0 \ne \phi$,  such that
\begin{equation}
\sum_{i\in \tilde{S}_1}r_i < I_{\mbox{\scriptsize \boldmath  $r$}} ( \mbox{\boldmath  $X$}_{i \in \tilde{S}_1} ; Y |\mbox{\boldmath  $X$}_{i \not\in S}).
\end{equation}
For the same reason, if $S\setminus \tilde{S}_1 \cap S_0 \ne \phi$, we can find a user subset $\tilde{S}_2 \subseteq S\setminus \tilde{S}_1  $, such that
\begin{equation}
\sum_{i\in \tilde{S}_2}r_i < I_{\mbox{\scriptsize \boldmath  $r$}} ( \mbox{\boldmath  $X$}_{i \in \tilde{S}_2} ; Y |\mbox{\boldmath  $X$}_{i \not\in S\setminus \tilde{S}_1}).
\end{equation}
In general, if $S\setminus \tilde{S}_1 \setminus \dots \setminus \tilde{S}_{j-1} \cap S_0 \ne \phi$, we can find $\tilde{S}_j \subseteq S\setminus \tilde{S}_1 \setminus \dots \setminus \tilde{S}_{j-1}$, such that
\begin{equation}
\sum_{i\in \tilde{S}_j}r_i < I_{\mbox{\scriptsize \boldmath  $r$}} ( \mbox{\boldmath  $X$}_{i \in \tilde{S}_j} ; Y |\mbox{\boldmath  $X$}_{i \not\in S\setminus \tilde{S}_1 \setminus \dots \setminus \tilde{S}_{j-1}}).
\label{RecursiveTildeS}
\end{equation}
This procedure can be carried out recursively until for some integer $j>0$, $S\setminus \tilde{S}_1 \setminus \dots \setminus \tilde{S}_{j} \cap S_0 = \phi$.

Consequently, define $\tilde{S}=\bigcup_{k=1}^j \tilde{S}_k $. Due to (\ref{RecursiveTildeS}), we have
\begin{eqnarray}
&& \sum_{i\in \tilde{S}}r_i < \sum_{k=1}^j I_{\mbox{\scriptsize \boldmath  $r$}} ( \mbox{\boldmath  $X$}_{i \in \tilde{S}_k} ; Y |\mbox{\boldmath  $X$}_{i \not\in S\setminus \tilde{S}_1 \setminus \dots \setminus \tilde{S}_{k-1}})     = I_{\mbox{\scriptsize \boldmath  $r$}} ( \mbox{\boldmath  $X$}_{i \in \tilde{S}} ; Y |\mbox{\boldmath  $X$}_{i \not\in S}).
\end{eqnarray}
Because $\tilde{S}\supseteq S\cap S_0$ can be found for any $S\subseteq \{1, \dots, K\}$, and $S \cap S_0 \ne \phi$, we conclude that $\mbox{\boldmath  $r$} \in \mbox{\boldmath  $R$}_{S_0}$.
\end{proof}

\subsection{Proof of Proposition \ref{Proposition1}}
\label{ProofProposition1}
\begin{proof}
Let $\mbox{\boldmath  $R$}$ be the achievable rate region given by Theorem \ref{Theorem2}. It is easy to show that any rate region $\tilde{\mbox{\boldmath  $R$}}\subseteq \mbox{\boldmath  $R$}$ is also asymptotically achievable. Therefore, we only need to show that, for all rate vector $\mbox{\boldmath  $r$}$ with $\sum_{i=1}^{K}\sqrt{r_k/n} < 1$, the following inequality holds for any user subset $S\subseteq \{1, \dots, K\}$:
\begin{equation}
\sum_{i\in S} r_i < I_{\mbox{\scriptsize \boldmath  $r$}} ( \mbox{\boldmath  $X$}_{i\in S} ; Y |\mbox{\boldmath  $X$}_{i \not\in S}).
\end{equation}

Note that since the channel output symbol $Y$ is a deterministic function of the channel input symbol vector $\mbox{\boldmath  $X$}$, we have $H_{\mbox{\scriptsize \boldmath  $r$}} (Y |\mbox{\boldmath  $X$})=0$. Consequently,
\begin{eqnarray}
&& I_{\mbox{\scriptsize \boldmath  $r$}} ( \mbox{\boldmath  $X$}_{i\in S} ; Y |\mbox{\boldmath  $X$}_{i \not\in S})=H_{\mbox{\scriptsize \boldmath  $r$}} (Y |\mbox{\boldmath  $X$}_{i \not\in S}) \nonumber \\
&& \quad \ge H_{\mbox{\scriptsize \boldmath  $r$}} (Y |\mbox{\boldmath  $X$}_{i \not\in S}= \mbox{\boldmath  $0$}) Pr\{\mbox{\boldmath  $X$}_{i \not\in S}= \mbox{\boldmath  $0$}\} \nonumber \\
&& \quad = H_{\mbox{\scriptsize \boldmath  $r$}} (Y |\mbox{\boldmath  $X$}_{i \not\in S}= \mbox{\boldmath  $0$}) \prod_{i\not\in S} (1-\sqrt{r_i/n}).
\end{eqnarray}
Since
\begin{eqnarray}
&& H_{\mbox{\scriptsize \boldmath  $r$}} (Y |\mbox{\boldmath  $X$}_{i \not\in S}= \mbox{\boldmath  $0$})\ge -\sum_{i=1}^{2^n} Pr\{Y=i\}\log_2Pr\{Y=i\} \nonumber \\
&& \quad \ge n Pr\{Y\not\in \{0, c\}\} \nonumber \\
&& \quad \ge n \sum_{i\in S} Pr\{X_i\ne 0, \mbox{\boldmath  $X$}_{k \in S\setminus \{i\}}= \mbox{\boldmath  $0$} \} \nonumber \\
&& \quad = n \sum_{i\in S} \sqrt{r_i/n} \prod_{k\not\in S\setminus \{i\}}(1-\sqrt{r_k/n}),
\end{eqnarray}
we indeed have
\begin{eqnarray}
&& I_{\mbox{\scriptsize \boldmath  $r$}} ( \mbox{\boldmath  $X$}_{i\in S} ; Y |\mbox{\boldmath  $X$}_{i \not\in S}) \ge n \sum_{i\in S} \sqrt{r_i/n} \prod_{k \ne i}\left(1-\sqrt{r_k/n}\right) \nonumber \\
&& \quad \ge n \sum_{i\in S} \sqrt{r_i/n}  \left(1-\sum_{k \ne i} \sqrt{r_k/n}\right) \nonumber \\
&& \quad \ge n \sum_{i\in S} \sqrt{r_i/n}  \left(\sqrt{r_i/n}+1-\sum_{k} \sqrt{r_k/n}\right) \nonumber \\
&& \quad > \sum_{i\in S} r_i,
\end{eqnarray}
where the last inequality follows from $\sum_{i=1}^{K}\sqrt{r_k/n} < 1$.
\end{proof}






\begin{thebibliography}{1}
\bibitem{ref Shannon48}
C. Shannon, {\em A Mathematical Theory of Communication,}
Bell System Technical Journal, Vol. 27, pp. 379-423, 623-656, July, October, 1948.

\bibitem{ref Cover05}
T. Cover and J. Thomas, {\em Elements of Information Theory,}
2nd Edition, Wiley Interscience, 2005.

\bibitem{ref Gallager65}
R. Gallager, {\em A Simple Derivation of The Coding Theorem and Some Applications,}
IEEE Transactions on Information Theory, Vol. 11, pp. 3-18, January 1965.

\bibitem{ref Telatar95}
I. Telatar and R. Gallager, {\em Combining Queueing Theory with Information Theory for Multiaccess,}
IEEE Journal on Selected Areas in Communications, Vol. 13, pp. 963–969, August 1995.

\bibitem{ref Berry02}
R. Berry and R. Gallager, {\em Communication over Fading Channels with Delay Constraints,}
IEEE Transactions on Information Theory, Vol. 48, pp. 1135–1149, May 2002.

\bibitem{ref Yeh02}
E. Yeh, {\em Delay-optimal Rate Allocation in Multiaccess Communications: A Cross-layer View,}
IEEE MMSP, St.Thomas, Virgin Islands, December 2002.

\bibitem{ref Zhao07}
Q. Zhao and B. Sadler, {\em A Survey of Dynamic Spectrum Access,}
IEEE Signal Processing Magazine, Vol. 24, pp. 79-89, May 2007.

\bibitem{ref Ephremides98}
A. Ephremides and B. Hajek, {\em Information Theory and Communication Networks: An Unconsummated Union,}
IEEE Transactions on Information Theory, Vol. 44, pp.2416-2434, October 1998.

\bibitem{ref Bertsekas92}
D. Bertsekas and R. Gallager, {\em Data Network,}
2nd Edition, Prentice Hall, 1992.

\bibitem{ref Luo06}
J. Luo and A. Ephremides, {\em On the Throughput, Capacity, and Stability Regions of Random Multiple Access,}
IEEE Transactions on Information Theory, Vol. 52, pp. 2593-2607, June 2006.

\bibitem{ref Shamai07}
S. Shamai, I. Teletar, and S. Verd\'{u}, {\em Fountain Capacity,}
IEEE Transactions on Information Theory, Vol. 53, pp. 4372-4376, November 2007.

\bibitem{ref Polyanskiy10}
Y. Polyanskiy, V. Poor, and S. Verd\'{u}, {\em Channel Coding Rate in The Finite Blocklength Regime,}
submitted to IEEE Transactions on Information Theory.

\bibitem{ref Forney68}
G. Forney, {\em Exponential Error Bounds for Erasure, List, and Decision Feedback Schemes,}
IEEE Transactions on Information Theory, Vol.14, pp. 206-220, March 1968.

\bibitem{ref Rao88}
R. Rao and A. Ephremides, {\em On The Stability of Interacting Queues in A Multiple-access System,}
IEEE Transactions on Information Theory, Vol. 34, pp. 918-930, September 1988.

\bibitem{ref Hui84}
J. Hui, {\em Multiple Accessing for The Collision Channel without Feedback,}
IEEE Journal on Selected Areas in Communications, Vol. SAC-2, pp. 575–582, July 1984.

\bibitem{ref Wang10}
Z. Wang and J. Luo, {\em Achievable Error Exponent of Channel Coding in Random Access Communication,}
IEEE ISIT, Austin, TX, June 2010.

\bibitem{ref Fano61}
R. Fano, {\em Transmission of Information,}
The M.I.T Press, and John Wiley \& Sons, Inc., New York, N.Y., 1961.

\bibitem{ref Berger77}
T. Berger, {\em Multiterminal Source Coding,}
In G. Longo Ed., The Information Theory Approach to Communications, Springer-Verlag, New York, 1977.

\bibitem{ref Csiszar81}
I. Csiszar and J. Korner, {\em Information Theory: Coding Theorems for Discrete Memoryless Systems,}
Academic Press, New York, 1981.

\end{thebibliography}
%




\end{document}